\documentclass[12pt]{article}
\usepackage{amsfonts}
\usepackage{amsmath}
\usepackage{amsbsy}

\font\grb=eurb10

\def\bphi{\hbox{\grb\char'047}\,}
\def\bpsi{\hbox{\grb\char'040}\,}
\def\bchi{\hbox{\grb\char'037}\,}
\def\bfi{\hbox{\grb\char'036}\,}

\def\I{\hbox{\bf I}}

\def\e{\hbox{\bf e}}
\def\f{\hbox{\bf f}}
\def\k{\hbox{\bf k}}
\def\l{\hbox{\bf l}}
\def\m{\hbox{\bf m}}

\def\G{\hbox{\bf G}}
\def\M{\hbox{\bf M}}
\def\N{\hbox{\bf N}}
\def\R{\hbox{\bf R}}
\def\U{\hbox{\bf U}}
\def\V{\hbox{\bf V}}
\def\W{\hbox{\bf W}}
\def\bigpsi{{\bf \Psi}}
\def\bigomega{{\bf \Omega}}

\begin{document}

\title{Collision of plane gravitational and electromagnetic waves in a Minkowski background: solution of the 
characteristic initial value problem}

\author{ G. A. Alekseev$^1$\thanks{E--mail: {\tt G. A. Alekseev@mi.ras.ru}}
and J. B. Griffiths$^2$\thanks{E--mail: {\tt J.B.Griffiths@Lboro.ac.uk}} \\ \\
$^1$ Steklov Mathematical Institute of Russian Academy of Sciences, \\Gubkina st. 8, 119991, Moscow, Russia\\ \\
$^2$ Department of Mathematical Sciences, Loughborough University, \\
Loughborough, Leics. LE11 3TU, U.K. \\ \\}

\date{\today}
\maketitle

\begin{abstract}
\noindent
We consider the collisions of plane gravitational and electromagnetic waves with distinct wavefronts and of arbitrary polarizations in a Minkowski background. We first present a new, completely geometric formulation of the characteristic initial value problem for solutions in the wave interaction region for which initial data are those associated with the approaching waves. We present also a general approach to the solution of this problem which enables us in principle to construct solutions in terms of the specified initial data. This is achieved by re-formulating the nonlinear dynamical equations for waves in terms of an associated linear problem on the spectral plane. A system of linear integral ``evolution'' equations which solve this spectral problem for specified initial data is constructed. It is then demonstrated explicitly how various colliding plane wave space-times can be constructed from given characteristic initial data.
\end{abstract}


\newpage
\tableofcontents

\newpage
\section{Introduction}

The collisions of plane gravitational or gravitational and electromagnetic waves in general relativity have been studied extensively over recent decades by many authors. These studies have revealed many interesting properties concerning the nonlinear interaction between such waves. These relate to both the physical and geometrical features of space-times with colliding plane waves. The great majority of these properties have been deduced from detailed studies of particular solutions of Einstein's equations, which were derived by some very special particular methods and tricks. However, a general mathematical solution of this problem, in which solutions in the wave interaction region are constructed from ``input data'' (the parameters characterizing the waves before their collision) has not been presented until now (although reported briefly in \cite{Alekseev-Griffiths:2001}).

In this paper, we present a general scheme for a solution of this problem for gravitational or gravitational and electromagnetic waves of arbitrary polarizations propagating with distinct wavefronts and colliding in a Minkowski background. This method is based on the well known integrability properties of the hyperbolic reductions of Einstein's vacuum and electrovacuum field equations with a two-dimensional Abelian space-time isometry group. Mathematical tools for the solution of the characteristic initial value problems for these equations have been developed in \cite{Alekseev:2001}. However, this approach has had to be adapted (with some necessary generalizations) for colliding plane waves in a Minkowski background.

In the remaining part of this introduction, we recall some elements of the theory of integrable hyperbolic reductions of the Einstein and
Einstein--Maxwell equations and give some references to preceding results. In the subsequent sections, we present a detail description of our new construction, the corresponding solution generation procedure, and we demonstrate how the procedure can be implemented. This demonstration includes the derivation of a previously unknown family of solutions.

Reviewing what is known about colliding plane waves, one could initially recall an appreciable number of known solutions of Einstein's equations
which can be interpreted as the outcomes of a collision of plane waves. A large number of such solutions and their physical and geometrical
interpretations can be found in~\cite{Griffiths:1991}. This list may be
extended by further publications appearing since 1991. In particular,
infinite hierarchies of exact vacuum and electro\-vacuum solutions with an arbitrary number of free parameters have recently been
found~\cite{Alekseev-Griffiths:2000}, and many of these are of the type that is appropriate for colliding plane waves. However, all of these colliding plane wave solutions have been found using an ``inverse'' approach in which a formal solution in the interaction region is directly obtained and the initial parameters of the corresponding approaching waves were calculated afterwards.

These solutions describe pairs of plane pure gravitational or mixed
gravitational and electromagnetic waves with distinct wavefronts which
approach each other from opposite spatial directions on the Minkowski
background. The problem of constructing the corresponding solution in the wave interaction region is a well formulated characteristic initial value problem. (See \cite{Griffiths:1991} and the sections below for a detailed formulation of this problem including the general properties of approaching waves, typical matching conditions and the structure of the
governing symmetry reduced field equations.)

In the important particular case in which the colliding waves are purely gravitational and possess constant and aligned polarizations, Einstein's equations in the interaction region can be reduced to a linear
Euler--Poisson--Darboux equation. In this case, the solution can be
constructed from the characteristic initial data using the generalized
version of Abel's transform \cite{Hauser-Ernst:1989},
\cite{Griffiths-SR:2002}. However, if the polarizations of the approaching gravitational waves are not constant and aligned, or in the presence of electromagnetic waves, the governing Einstein or Einstein--Maxwell field equations are essentially nonlinear and this makes the problem much more complicated.

Fortunately, these equations (which can be represented conveniently in the form of hyperbolic Ernst equations) have been found to be completely integrable. In fact, the integrability of the hyperbolic vacuum and
electrovacuum Ernst equations facilitates a number of solution generating techniques such as the generation of vacuum \cite{Belinskii-Zakharov:1978} and electrovacuum \cite{Alekseev:1980} solitons, B\"acklund or the symmetry transformations \cite{Harrison:1978,Neugebauer:1979,Harrison:1983}. However, no techniques have been developed which provide algorithms for explicitly generating solutions from initial data. Nonetheless, this integrability gives rise also to some powerful methods for a general analysis of the governing field equations. In particular, Hauser and Ernst analyzed the characteristic initial value problem for the vacuum hyperbolic Ernst equation \cite{Hauser-Ernst:1990-1991}. They generalized their group-theoretical approach (which had been developed earlier for stationary axisymmetric fields) and constructed a homogeneous Hilbert problem with corresponding matrix linear integral equations. Many global aspects of the characteristic initial value for vacuum fields, including the existence and uniqueness of solutions and a detailed proof of the Geroch conjecture, have been elaborated in this way \cite{Hauser-Ernst:2001}. However, this approach does not immediately lead to any effective methods for the solution of the corresponding characteristic initial value problem. 

Other schemes for the solution of characteristic initial value problems for the vacuum and electrovavuum Ernst equations \cite{Alekseev:1993} or for the  $\sigma$-model form of the vacuum equations \cite{Fokas-Sung-Tsoubelis:1999} have also been considered. However, in these cases, characteristic initial value problems were analyzed only for initial data which are regular functions of the geometrically defined coordinates (\ref{xietadef}) below. As will be discussed below, this restriction does not cover the collisions of plane waves with distinct wavefronts in a Minkowski background. In such cases, physically relevant initial data, being regular on the wavefronts in the local frame, must possess some singularities as functions of the coordinates adopted.

The solution of the characteristic initial value problem for colliding plane waves that will be presented here originates from another approach to the analysis of the structure of integrable reductions of the Einstein and Einstein--Maxwell field equations that was developed in
\cite{Alekseev:1985}--\cite{Alekseev:2000}. In this approach, every
solution can be characterized by a set of functions of an auxiliary
(spectral) parameter. These functions are interpreted as the monodromy data on the spectral plane of the fundamental solution of an associated linear system with a spectral parameter. These monodromy data are nonevolving (i.e. coordinate independent) and, generally, can be chosen arbitrarily or specified in accordance with the properties of the solution being sought. In particular, these data can be determined (at least in principle) from the initial or boundary data. In this scheme, the solution of the initial or boundary value problem is determined by the solution of some linear singular integral equations whose scalar kernel is constructed using these (specified) monodromy data. In this original form of the ``monodromy transform'' approach, a conjecture of local analyticity of solutions near the point of normalization was used very essentially. However, for waves with distinct wavefronts, the solutions of the linear equations are normalized at the point at which the waves collide and, for this situation, these are nonanalytic at this point. This violation of analyticity at the point of normalization gives rise to some difficulties with the construction of the master singular integral equation.

Our resolution of this difficulty is achieved by the introduction of some
``dressing'' or ``scattering'' matrices (see \cite{Alekseev:2001}) into the structure of the fundamental solution of the associated linear system. The problem can then be overcome because, even in the nonanalytical case, these matrices possess much more convenient analytical properties. For locally analytic solutions, this construction is completely equivalent to the previous approach. However, the new construction introduces some interesting features. In particular, the monodromy data which characterize the scattering matrices are not conserved quantities. They evolve, and are therefore referred to as ``dynamical monodromy data''. Moreover, the evolution of these data is determined completely by the behaviour of fields on the characteristics which cross at the point of normalization. Again, the usual nonlinear partial differential form of the field equations are replaced by a pair of linear integral equations which determine the solution uniquely. With this new construction, the scalar kernels of these new quasi-Fredholm equations are built from this new kind of monodromy data. These linear integral equations then essentially possess a different, simpler structure. Moreover, they are much better adapted to the context of
a characteristic initial value problem.

\section{Space-time geometry for colliding plane waves}

\paragraph*{Symmetry conjecture.}
In the situation under consideration, each of the approaching plane waves
possesses two commuting space-like Killing vector fields with noncompact
orbits, and we expect this symmetry to be global and retained during the
collision and subsequent interaction of the waves. Thus, throughout the
space-time, the metric and all nonmetric field components can be considered
locally to be functions of two null coordinates, say $u$ and $v$, while the
two other space-like coordinates, say $y$ and $z$ (which we assume to be
defined globally) are completely ignorable. An additional simplification of
the metric and electromagnetic field components arises if we extend the
conjecture of $y$- and $z$- independence from the metric and Maxwell tensor
components to include their related potentials.

\paragraph*{Conjectured structure of Einstein--Maxwell fields.}
For both colliding plane gravitational waves in vacuum and a coupled
system of colliding plane gravitational and electromagnetic waves, the
components of the metric and the 1-form of the electromagnetic vector
potential can be considered globally in the forms
     \begin{equation}
     \label{Metric}
ds^2=g_{\mu\nu}dx^\mu dx^\nu+g_{ab} dx^a dx^b,\qquad
\underline{\bf A}=A_b dx^b
     \end{equation}
     where  $\mu,\nu,\ldots=0,1$ with $x^0,x^1=u,v$, and $a,b,\ldots=2,3$
with $x^2,x^3=y,z$. The orbit space metric $g_{\mu\nu}$, the metric 
on the orbits $g_{ab}$ and the components $A_b$ of the vector 
electromagnetic potential are functions of $x^\mu$ only.

\paragraph*{General and geometrically defined coordinates.}
As is well known, the local coordinates $x^\mu$ can be specified to reduce
$g_{\mu\nu}$ to a conformally flat form:
     \begin{equation}
     \label{ConfCoords}
g_{\mu\nu}dx^\mu dx^\nu=f(u,v) du dv.
     \end{equation}
  It is clear that these null coordinates are defined up to an arbitrary
transformation of the form
     \begin{equation}
     \label{uv2uv}
\widetilde{u}=\widetilde{u}(u)\quad\mbox{and}\quad
\widetilde{v}=\widetilde{v}(v).
     \end{equation}

For the analysis of the reduced Einstein--Maxwell equations and their 
integrability properties in the wave interaction region, it is 
convenient to use also some geometrically defined null coordinates 
$\xi$ and $\eta$ which can be introduced as follows:
     \begin{equation}
     \label{xietadef}
     \left\{\begin{array}{l}
     \xi=\beta+\alpha\\
     \eta=\beta-\alpha
     \end{array}\right.
     \quad\mbox{where}\qquad     
     \alpha=\sqrt{\det\Vert g_{ab}\Vert}\quad
\mbox{and}\quad \left\{\begin{array}{l}\partial_u \beta=\partial_u\alpha\\
\partial_v \beta=-\partial_v\alpha.\end{array}\right.
    \end{equation}
The function $\alpha(u,v)$ is the area measure on the orbits of
the isometry group and the function $\beta(u,v)$ is then determined (up to an additive real constant).  (It may be recalled that the Einstein--Maxwell equations imply that the function
$\alpha(u,v)$ is a ``harmonic'' function in a sense that it satisfies the
equation $\partial_u\partial_v\alpha=0$ which also provides the integrability condition for the above $\beta$-equations to be satisfied.)  

\paragraph*{Parametrization of the metric components.}
It is convenient to parametrize the metric on the orbits by three scalar functions $\alpha(u,v)>0$, $H(u,v)>0$ and $\Omega(u,v)$ as
     \begin{equation}
     \label{OrbitMetric}
     g_{ab}=-\begin{pmatrix}H& H\Omega\\
H\Omega & H\Omega^2+\displaystyle{\frac{\alpha^2} H}\end{pmatrix}.
     \end{equation}

\paragraph*{Newman--Penrose tetrad and scalars.}
With the metric (\ref{Metric}) it is convenient to adopt the
Newman--Penrose null tetrad whose vector components with respect to coordinates $(u,v,y,z)$ are:
     $$\begin{array}{lcclccl}
{\bf l}_j=\displaystyle{\sqrt{\frac f 2}}\{1,0,0,0\}&&&
{\bf n}_j=\displaystyle{\sqrt{\frac f 2}}\{0,1,0,0\}&&&
{\bf m}_j=\displaystyle{\sqrt{\frac H 2}}\{0,0,1,\Omega+\dfrac {i\alpha} H\}\\[2ex]
{\bf l}^j=\displaystyle{\frac 1{\sqrt{2 f}}}\{0,1,0,0\}&&&
{\bf n}^j=\displaystyle{\frac 1{\sqrt{2 f}}}\{1,0,0,0\}&&&
{\bf m}^j=\displaystyle{\frac i\alpha}\displaystyle{\sqrt{\dfrac H
2}}\{0,0,\Omega+\frac {i\alpha} H,-1\}
     \end{array}$$
    where $j=0,1,2,3$. The expressions for the standard projections of the self-dual parts of the Weyl tensor ($\Psi_0$, $\Psi_1$, $\Psi_2$, $\Psi_3$, $\Psi_4$) and of the Maxwell tensor ($\phi_0$, $\phi_1$, $\phi_2$) then take the forms
\begin{equation}\label{NPscalars}
\begin{array}{lcccl}
\Psi_0=-\dfrac i\alpha (\partial_v+i\alpha^{-1} H\partial_v\Omega)
\left[\dfrac H f\partial_v(\Omega+\dfrac {i\alpha} H)\right] &&&&
\phi_0=-\displaystyle{\dfrac 2{\sqrt{f H}}} \partial_v\Phi\\
\Psi_1=0 &&&& \phi_1=0\\
\Psi_2=\dfrac{H^2}{2 f\alpha^2}\partial_v(\Omega+\dfrac {i\alpha} H)
\partial_u(\Omega-\frac {i\alpha}
H)-\dfrac{\partial_u\alpha\partial_v\alpha}{2 f
\alpha^2} &&&&
\phi_2=\displaystyle{\dfrac 2{\sqrt{f H}}}\partial_u\Phi\\
\Psi_3=0 &&&&\\
\Psi_4=\dfrac i\alpha (\partial_u-i\alpha^{-1} H\partial_u\Omega)
\left[\dfrac H f\partial_u(\Omega-\dfrac {i\alpha} H)\right]&&&&
     \end{array}
     \end{equation}
     where $\Phi$ is the Ernst scalar electromagnetic potential to be defined
below. (This potential coincides with the $y$-component ($\Phi\equiv\Phi_y$)
of the complex self-dual electromagnetic vector potential
$\Phi_i=\{0,0,\Phi_y,\Phi_z\}$.)

\paragraph*{The symmetry reduced Einstein--Maxwell equations.}

It is well known that, for these space-times, the nontrivial part of the
Einstein--Maxwell equations decouples into two parts. One of these contains the constraint equations which can be considered as determining the conformal factor $f(u,v)$ (up to a multiplicative constant) in terms of the other functions:
     \begin{equation}
    \label{CFactor}
\left\{\begin{array}{l}
\displaystyle{\frac {f_u}f}=\displaystyle{\frac{\alpha_{uu}}
{\alpha_u}}-\displaystyle{\frac {H_u}H}+
\displaystyle{\frac \alpha {2\alpha_u}}\left[\displaystyle{\frac
{\left\vert{\cal
E}_u+2\overline{\Phi}\Phi_u\right\vert^2}
{H^2}}+\displaystyle{\frac 4 H} \vert\Phi_u\vert^2\right]\\[3ex]
\displaystyle{\frac {f_v} f}=\displaystyle{\frac {\alpha_{vv}}
{\alpha_v}}-\displaystyle{\frac {H_v} H}+\displaystyle{\frac \alpha
{2\alpha_v}}
\left[\displaystyle{\frac {\left\vert{\cal
E}_v+2\overline{\Phi}\Phi_v\right\vert^2}
{H^2}}+\displaystyle{\frac 4 H}\vert\Phi_v\vert^2\right]
     \end{array}\right.
     \end{equation}
where all suffices mean the derivatives, and the right hand sides
are expressed in terms of the complex Ernst potentials ${\cal E}(u,v)$ and $\Phi(u,v)$ which are determined by the relations:
     \begin{equation}
     \label{EFPotentials}
\left\{\begin{array}{rl}
\mbox{Re}\,{\cal E}=&-H-\Phi\overline{\Phi}\\
\partial_u(\mbox{Im}\,{\cal E})=&\alpha^{-1} H^2
\partial_u\Omega+ i(\overline{\Phi}\partial_u\Phi-\Phi
\partial_u\overline{\Phi})\\
\partial_v(\mbox{Im}\,{\cal E})=&-\alpha^{-1} H^2
\partial_v\Omega+ i(\overline{\Phi}\partial_v\Phi-\Phi
\partial_v\overline{\Phi})
     \end{array}\right.\quad
     \left\{\begin{array}{rl}
\mbox{Re}\,\Phi=&A_y\\
\partial_u(\mbox{Im}\,\Phi)=&-\alpha^{-1} H(\partial_u A_z-\Omega
\partial_u A_y)\\
\partial_v(\mbox{Im}\,\Phi)=&\alpha^{-1} H(\partial_v A_z-\Omega
\partial_v A_y)
     \end{array}\right.
     \end{equation}
  where $A_y$, $A_z$ are the nonzero components of a real electromagnetic
vector potential.

The remaining part of the reduced electrovacuum Einstein--Maxwell
equations are the dynamical equations for the metric functions
$\alpha(u,v)$, $H(u,v)$ and $\Omega(u,v)$. It is very convenient to
present these in the form of the hyperbolic Ernst equations for the Ernst
potentials:
     \begin{equation}
     \label{ErnstEqs}
\left\{\begin{array}{l} (\mbox{Re}\,{\cal
E}+\Phi\overline{\Phi})\left(2{\cal E}_{uv}+ \displaystyle{\frac
{\alpha_u}\alpha}\,{\cal
E}_v+\displaystyle{\frac{\alpha_v}\alpha}\,{\cal
E}_u \right)-\left({\cal E}_u+2\overline{\Phi}\Phi_u\right) \,{\cal
E}_v-\left({\cal E}_v+2\overline{\Phi}\Phi_v\right)
\,{\cal E}_u=0\\[1ex]
(\mbox{Re}\,{\cal E}+\Phi\overline{\Phi})\left(2\Phi_{uv}+
\displaystyle{\frac{\alpha_u}\alpha}\Phi_v+
\displaystyle{\frac{\alpha_v}\alpha}\Phi_u\right)-\left({\cal E}_u+2\overline{\Phi}\Phi_u\right)\Phi_v-
\left({\cal E}_v+2\overline{\Phi}\Phi_v\right)\Phi_u=0\\[1ex]
\alpha_{uv}=0
\end{array}\right.
     \end{equation}

These dynamical equations play the primary role in studies of colliding
plane waves because they govern the nonlinear processes of the interaction between the waves. They also provide the integrability conditions for the constraint equations (\ref{CFactor}) and the compatibility of the relations (\ref{EFPotentials}) which relate all the metric functions and other field variables with the solutions of the Ernst equations (\ref{ErnstEqs}). The constraint equations  (\ref{CFactor})  determine the remaining metric component -- the conformal factor $f(u,v)$.

\paragraph*{Matching conditions at the wavefronts.}

We say that a plane wave with a distinct wavefront, say $u=0$ or $v=0$,
propagates through a given background if any wave-like solution of
(\ref{ErnstEqs}) defined for $u\ge 0$ or $v\ge 0$ respectively matches
appropriately on the corresponding wavefront to the given background
solution which is defined in the region $u\le 0$ or $v\le 0$ respectively.

For colliding plane waves, it is well known that when considering the
junction conditions across the null hypersurfaces which correspond to the
wavefronts, the familiar Lichnerowicz conditions, which require that there should exist a coordinate system in which the components of the metric and electromagnetic potential are at least of class $C^1$ on the wavefront, should be relaxed  in favour of the O'Brien--Synge conditions. These admit also some other types of waves, such as impulsive gravitational waves or gravitational and electromagnetic shock waves.
For the class of metrics defined above, these conditions imply that across the wavefront

(i)  the function $\alpha$ is of the class $C^1$

(ii) the Ernst potentials ${\cal E}$ and $\Phi$ are continuous

(iii) the conformal factor $f$ is continuous

It is important to note here (see below for more details) that the
condition (iii), through the relations (\ref{CFactor}), may impose some
further restrictions (additional to the conditions (i) and (ii)) on the
behaviour of function $\alpha$ and the Ernst potentials near the wavefront.

\section{Plane waves in a Minkowski background}
\label{SingleWave}

\paragraph*{The Minkowski background.}
Within the class of metrics (\ref{Metric}), the Minkowski space-time
can be represented in the form
     \begin{equation}
     \label{MinkowskiA}
     ds^2=du dv -dy^2-dz^2
     \end{equation}
or, in the notations introduced above, by the values
     \begin{equation}
     \label{MinkowskiB}
     \alpha=1,\qquad{\cal E}=-1,\qquad\Phi=0,\qquad f=1
     \end{equation}
     (and therefore, $H=1$ and $\Omega=0$).
Everywhere below, we shall consider this space-time as the
background for the waves in the sense that was formulated in the 
previous section.

\paragraph*{Travelling waves.}
It is well known that the Ernst equations (\ref{ErnstEqs}) admit  a wide
class of plane wave solutions which can also be called ``travelling waves".
For these solutions all potentials  depend on one null coordinate $u$ or $v$
only and, therefore, correspond to plane waves travelling without any
evolution in the positive  or negative  $x$-direction respectively. To
distinguish these similar solutions, we shall call the waves depending on
$u$, and travelling therefore in the positive $x$-direction, the ``left
waves", and the waves depending on $v$ and travelling in the negative
$x$-direction the ``right waves".

\paragraph*{The left waves.}
For these waves all metric components and potentials depend only on $u$. It
is very convenient to use the coordinate freedom $u\to\widehat{u}(u)$ to
specify the conformal factor $f(u)=1$. Geometrically this means that the
coordinate $u$ is chosen  to be an affine
parameter on the null geodesics $v=\mbox{const}$, $y=\mbox{const}$,
$z=\mbox{const}$ which cross the wave. This class of waves can be
described completely by the Ernst potentials
  \begin{equation}\label{LeftWaveA}
\{{\cal E}(u),\,\, \Phi(u)\}
  \end{equation}
  which are functions of the affine parameter $u$ and which  should satisfy
only the signature condition
  \begin{equation}\label{LeftWaveB}
H(u)\equiv -\mbox{Re}\,{\cal E}(u)-\Phi(u)
\overline{\Phi}(u)> 0.
  \end{equation}
  The corresponding function $\alpha(u)$ is determined for $u>0$  as the solution of a linear differential equation which follows from the first of the constraint equations (\ref{CFactor}) for the case $f(u)=1$:
  \begin{equation}
  \label{LeftWaveC}
\alpha_{uu}-\displaystyle{\frac {H_u}H}\alpha_u+
\displaystyle{\frac 12}\left[\displaystyle{\frac
{\left\vert{\cal
E}_u+2\overline{\Phi}\Phi_u\right\vert^2}
{H^2}}+\displaystyle{\frac 4 H}\vert\Phi_u\vert^2\right]\alpha=0.
  \end{equation}
with the initial data at $u=0$ which follows from the matching conditions. 

\bigskip
\setlength{\unitlength}{0.85mm}
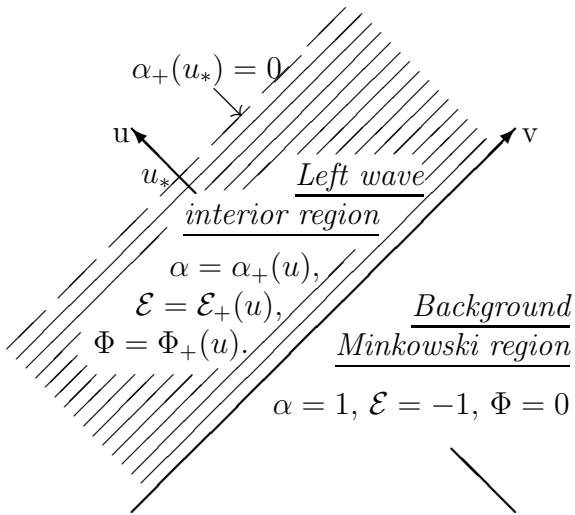
\begin{figure}[h]
\begin{center}
\begin{picture}(90,73)
\linethickness{0.1pt}
\put(12,58){\mbox{u}}
\put(76,58){\mbox{v}}
\put(16.5,4.5){\line(1,1){54}}
\put(15,6){\line(1,1){35}}
\put(13.5,7.5){\line(1,1){15}}
\put(12,9){\line(1,1){12.5}}
\put(10.5,10.5){\line(1,1){11}}
\put(9,12){\line(1,1){9.5}}
\put(7.5,13.5){\line(1,1){8}}
\put(6,15){\line(1,1){6.5}}
\put(4.5,16.5){\line(1,1){5}}
\put(3,18){\line(1,1){5}}
\put(1.5,19.5){\line(1,1){5}}
\put(0,21){\line(1,1){7}}
\put(-1.5,22.5){\line(1,1){22}}
\put(-3,24){\line(1,1){54}}

\put(-4.5,25.5){\line(1,1){6}}
\put(3.5,33.5){\line(1,1){6}}
\put(11.5,41.5){\line(1,1){6}}
\put(19.5,49.5){\line(1,1){6}}
\put(27.5,57.5){\line(1,1){6}}
\put(35.5,65.5){\line(1,1){6}}
\put(42.5,73){\line(1,1){6}}

\put(15,68){$\alpha_+(u_\ast)=0$}
\put(27.5,63){$\searrow$}
\put(16.5,51.5){$u_\ast$}

\put(69,60){\line(-1,-1){8}}
\put(67.5,61.5){\line(-1,-1){7}}
\put(66,63){\line(-1,-1){7}}
\put(64.5,64.5){\line(-1,-1){8.5}}
\put(63,66){\line(-1,-1){10}}
\put(61.5,67.5){\line(-1,-1){11.5}}
\put(60,69){\line(-1,-1){13}}
\put(58.5,70.5){\line(-1,-1){14.5}}
\put(57,72){\line(-1,-1){16}}
\put(55.5,73.5){\line(-1,-1){23}}
\put(54,75){\line(-1,-1){24.5}}
\put(52.5,76.5){\line(-1,-1){26}}
\thicklines
\put(15,0){\vector(1,1){60}}
\put(25,50){\vector(-1,1){10}}
\put(75,0){\line(-1,1){10}}

\put(21,38){\mbox{$\begin{array}{l}
\hskip8.2ex \underline{\hbox{\em Left wave}}\\
\underline{\hbox{\em interior region}}\\[1ex]
\hskip-1ex \alpha=\alpha_+(u),\\
\hskip-3.5ex {\cal E}={\cal E}_+(u),\\
\hskip-6.6ex \Phi=\Phi_+(u).
\end{array}$}}

\put(35,23){$\begin{array}{r}
\underline{\hbox{\em Background}}\\
\underline{\hbox{\em Minkowski region}}\\[1.5ex]
\alpha=1,\,{\cal E}=-1,\,\Phi=0
\end{array}$}

\end{picture}
\end{center}
\caption{\small The left wave propagating on the Minkowski background. The coordinate $u$ is chosen to be an affine parameter on the geodesics
$v=\mbox{const}$. In this case, $f=1$ everywhere.
\label{fig:LeftWave}}
\end{figure}

These waves for $u\ge0$ are combined with a Minkowski background for
$u<0$ as illustrated in Fig.(1). These two regions are joined on the null hypersurface (wavefront) $u=0$ where all metric functions and the Ernst potentials (\ref{LeftWaveA}) defined for $u\ge 0$ (the wave interior region) should be matched with their Minkowski values
(\ref{MinkowskiB}).

The matching conditions (i)--(iii) of the previous section imply the
following conditions on the wavefront $u=0$
  \begin{equation}
  \label{LeftWaveD}
{\cal E}(0)=-1,\quad \Phi(0)=0,\quad\alpha(0)=1,\quad\alpha'(0)=0
  \end{equation}
(together with the conditions $H(0)=1$ and $\Omega(0)=0$). The last two
conditions in (\ref{LeftWaveD}) provide initial conditions for the equation
(\ref{LeftWaveC}) for $\alpha(u)$ which is then determined uniquely. 
It follows from the structure of equation (\ref{LeftWaveC}) for $H(u)>0$ that, for any nonconstant data ${\cal E}(u)$,
$\Phi(u)$, the corresponding function $\alpha(u)$ decreases monotonically
from its initial value $\alpha(0)=1$ with $\alpha'(0)=0$, when $u$ runs from
$0$ to some finite critical  value $u_\ast$ where $\alpha(u_\ast)$ vanishes.
Since $\alpha$ is a measure of the area of two-dimensional sections of the
null geodesic tubes, its decrease is a manifestation of the well known
general focusing property of gravitational fields. The surface $u=u_\ast$
corresponds to the presence of a caustic, and represents a boundary of the
space-time being considered.

\paragraph*{The right waves.}
Solutions in which the metric components and the Ernst potentials are
functions of the null coordinate $v$ only are exactly equivalent to those
described above, except that they propagate in the opposite, negative
$x$-direction. In an exactly equivalent way to that described above, we
impose the coordinate condition $f(v)=1$ so that $v$ is an affine parameter and any solution is then described by the corresponding Ernst potentials $\{{\cal E}(v),\,\, \Phi(v)\}$ and the function $\alpha(v)>0$ satisfies an equation equivalent to (\ref{LeftWaveC}). Again, we introduce a wave of this type for $v\ge0$ by matching it with a Minkowski region for $v<0$ exactly as above.

\paragraph*{Other coordinates.}
Sometimes the wave solution simplifies considerably if we relax our choice of the null coordinate $u$ (or $v$) as the affine parameter on the null geodesics crossing the travelling wave interior region.

In practice, it is often convenient to avoid having to solve the equation
(\ref{LeftWaveC}) for $\alpha$. Instead, the freedom $u\to\tilde{u}(u)$ can be used to specify the form of $\alpha(u)$. However, this introduces a non-constant conformal function $f(u)$ which, for any wave functions ${\cal E}(u)$, $\Phi(u)$, can be calculated in quadratures from the constraint equations (\ref{CFactor}). Of course, the choice of the function $\alpha$ is restricted by the condition that it and its first derivative must be continuous on the wavefront and monotonically decreasing through the wave region. In addition, the choice of the Ernst potentials ${\cal E}(u)$ and $\Phi(u)$ should be also restricted by the condition for the regularity of the space-time geometry and the electromagnetic field components near the wavefront, and this imposes further restrictions on the character of $\alpha$ at the wavefront. This alternative approach will not generally be adopted in the remainder of this paper.

\section{Characteristic initial value problem for colliding plane waves}
\label{CharIVP}

Now we consider the situation in which two shock waves exist in the same
space-time, are initially separated by some Minkowski region, and approach
each other from opposite spatial directions. The left wave comes from the
negative $x$-direction and has the wavefront $u=0$, while the right wave
comes from the positive $x$-direction and has the wavefront $v=0$.

Before their collision, their interior regions ($u>0$, $v<0$ for the left wave
and  $v>0$, $u<0$ for the right wave) are described completely by the
Ernst potentials expressed as functions of the affine parameters on the
null geodesics crossing these regions. These are represented by the
functions ${\cal E}_+(u)$, $\Phi_+(u)$ for $u>0$ and ${\cal E}_-(v)$,
$\Phi_-(v)$ for $v>0$ for the left and right waves respectively. (For
convenience, we denote functions associated with the left and right waves
using the subscripts $+$ and $-$ respectively.) As discussed in the previous
section, the function $\alpha$ which appears in the dynamical equations as
an additional dynamical variable, is not an independent function. Rather,
in the approaching waves, the functions $\alpha_+(u)$ and $\alpha_-(v)$ are
determined uniquely as solutions of the corresponding constraint equations
with initial data determined by the matching conditions on their wavefronts.

A collision of these waves at the point $u=0$, $v=0$ gives rise to the
existence in space-time of a wave interaction region $(u>0,v>0)$. As in
the approaching wave regions, this region also possesses a boundary which is
determined by the condition $\alpha(u,v)=0$. However, unlike the boundaries
$u=u_\ast$ and $v=v_\ast$ of the approaching wave regions, this boundary is
generally a curvature singularity (see \cite{Griffiths:1991} for more
details). The complete space-time is illustrated in Fig.(3).

\bigskip\bigskip
\setlength{\unitlength}{0.9mm}
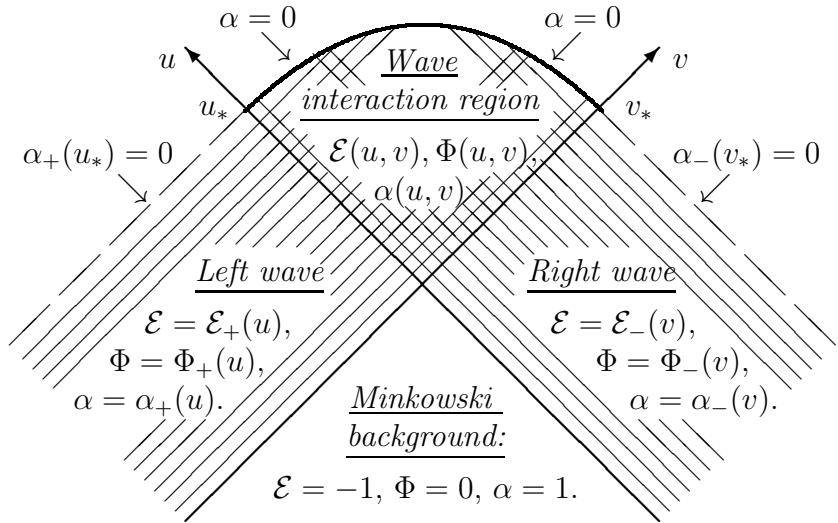
\begin{figure}[h]
\begin{center}
\hskip-6ex
\begin{picture}(90,70)
\linethickness{0.1pt}


\put(26,10){$\begin{array}{c}
\underline{\hbox{\em Minkowski }}\\
\underline{\hbox{\em background:}}\\[1ex]
{\cal E}=-1,\,\Phi=0,\,\alpha=1.
\end{array}$}

\put(-3.5,26){\mbox{$\begin{array}{l}
\hskip8.9ex \underline{\hbox{\em Left wave}}\\[1ex]
\hskip5.2ex{\cal E}={\cal E}_+(u),\\
\hskip2.6ex\Phi=\Phi_+(u),\\
\alpha=\alpha_+(u).
\end{array}$}}

\put(-11,26){\line(1,1){6}}
\put(-3.6,33.4){\line(1,1){6}}
\put(3.8,40.8){\line(1,1){6}}
\put(11.2,48.2){\line(1,1){6}}
\put(18.6,55.6){\line(1,1){6}}

\put(-9.0,24.0){\line(1,1){46}}
\put(-7.5,22.5){\line(1,1){38.0}}
\put(34.5,64.5){\line(1,1){7.3}}

\put(-6.0,21.0){\line(1,1){37.0}}
\put(38.0,65.0){\line(1,1){8.0}}

\put(-4.5,19.7){\line(1,1){37.5}}
\put(19.5,41.0){\line(1,1){14.5}}
\put(22.5,41.0){\line(1,1){12}}
\put(25.0,40.5){\line(1,1){11}}
\put(28.5,40.5){\line(1,1){10.5}}
\put(31.5,40.5){\line(1,1){9.5}}
\put(33.5,39.5){\line(1,1){8}}
\put(58.5,64.5){\line(1,1){5.1}}

\put(35.0,38.0){\line(1,1){9}}
\put(61.0,64.0){\line(1,1){5.1}}

\put(36.5,36.5){\line(1,1){9}}
\put(36.5,33.5){\line(1,1){12.5}}
\put(10.5,4.5){\line(1,1){42}}
\put(12.0,3.0){\line(1,1){48}}
\put(13.5,1.5){\line(1,1){60.8}}

\put(9,6){\line(1,1){9}}
\put(7.5,7.5){\line(1,1){7.5}}
\put(6,9){\line(1,1){6}}

\put(30,57){$\begin{array}{c}
\underline{\hbox{\em Wave}}\\
\underline{\hbox{\em interaction region}}\\[1ex]
\hskip2ex{\cal E}(u,v),\Phi(u,v),\\
\alpha(u,v)
\end{array}$}
\linethickness{1.0pt}
\qbezier(23.6,60.6)(50,86.2)(76.4,60.6)
\linethickness{0.2pt}

\put(61.5,26){\mbox{$\begin{array}{l}
\hskip1ex\underline{\hbox{\em Right wave}}\\[1ex]
\hskip2.7ex {\cal E}={\cal E}_-(v),\\
\hskip6ex\Phi=\Phi_-(v),\\
\hskip8.5ex \alpha=\alpha_-(v).
\end{array}$}}

\put(111,26){\line(-1,1){6}}
\put(103.6,33.4){\line(-1,1){6}}
\put(96.2,40.8){\line(-1,1){6}}
\put(88.8,48.2){\line(-1,1){6}}
\put(81.4,55.6){\line(-1,1){6}}

\put(87,53){$\alpha_-(v_\ast)=0$}
\put(91,48){$\swarrow$}
\put(-9,53){$\alpha_+(u_\ast)=0$}
\put(4.8,48){$\searrow$}

\put(109.0,24.0){\line(-1,1){46}}
\put(107.5,22.5){\line(-1,1){38}}
\put(65.5,64.5){\line(-1,1){7.8}}

\put(106.0,21.0){\line(-1,1){37}}
\put(62.0,65.0){\line(-1,1){8}}
\put(104.5,19.5){\line(-1,1){37.5}}
\put(80.5,40.5){\line(-1,1){13.0}}
\put(77.5,40.5){\line(-1,1){12}}
\put(74.5,40.5){\line(-1,1){11.0}}
\put(71.5,40.5){\line(-1,1){11}}
\put(68.5,40.5){\line(-1,1){10}}
\put(66.5,39.5){\line(-1,1){10}}
\put(41.5,64.5){\line(-1,1){5.1}}

\put(65.0,38.0){\line(-1,1){9}}
\put(39.0,64.0){\line(-1,1){5.1}}

\put(63.5,36.5){\line(-1,1){9}}
\put(63.5,33.5){\line(-1,1){12.5}}
\put(89.5,4.5){\line(-1,1){42}}
\put(88.0,3.0){\line(-1,1){44}}
\put(86.5,1.5){\line(-1,1){60.8}}

\put(91,6){\line(-1,1){9}}
\put(92.5,7.5){\line(-1,1){7.5}}
\put(94.5,9){\line(-1,1){6}}

\thicklines
\put(15,0){\vector(1,1){70}}
\put(87,67){$v$}
\put(80,60){$v_\ast$}
\put(85,0){\vector(-1,1){70}}
\put(11,67){$u$}
\put(17,60){$u_\ast$}

\put(20,73){$\alpha=0$}
\put(26,69){$\searrow$}
\put(68,73){$\alpha=0$}
\put(69,69){$\swarrow$}

\end{picture}
\end{center}
\caption{\small The collision of travelling plane waves propagating into a
Minkowski background. The coordinates $u$ and $v$ in the wave regions before
their collision are chosen to be affine parameters along the geodesics
$v=\mbox{const}$ and $u=\mbox{const}$ respectively. The conformal factors in
these regions are $f_+(u)=1$ and $f_-(v)=1$ respectively. However, in the
wave interaction region, $f(u,v)\ne 1$ and
$\alpha(u,v)=\alpha_+(u)+\alpha_-(v)-1$.
\label{fig:Collision}}
\end{figure}

The problem of the construction of the solution in the wave interaction region
for a given pair of approaching travelling plane waves is clearly a
characteristic initial value problem. The matching conditions applied on the
boundaries $(u=0,v>0)$ and $(v=0,u>0)$ of the wave interaction region imply
the continuity of the function $\alpha$ together with its first derivatives
and continuity of the Ernst potentials on these characteristics. These
conditions provide all the necessary characteristic initial values for these
functions.

First of all, it is easy to see that the characteristic initial value
problem for the function $\alpha(u,v)$ separates from that for the Ernst
potentials and its solution is trivial:
   \begin{equation}
     \label{Alpha}
     \alpha(u,v)=\alpha_+(u)+\alpha_-(v)-1,
     \end{equation}
  because the matching conditions for the approaching waves with the Minkowski
background have been satisfied, and the functions $\alpha_+(u)$ and
$\alpha_-(v)$ possess the properties
  \begin{equation}
     \label{Alpha0}
\alpha_+(0)=\alpha_-(0)=1,\qquad \alpha_+'(0)=\alpha_-'(0)=0.
  \end{equation}
  In particular, at the point of collision we have the
``normalization'' $\alpha(0,0)=1$.

We also have the characteristic initial data for the Ernst potentials in
the interaction region:
    \begin{equation}
    \label{BoundaryData}
    \begin{matrix}
{\cal E}(u,0)={\cal E}_+(u), \\[2pt]
\Phi(u,0)=\Phi_+(u),
    \end{matrix} \qquad
    \begin{matrix}
{\cal E}(0,v)={\cal E}_-(v), \\[2pt]
\Phi(0,v)=\Phi_-(v).
    \end{matrix}
    \end{equation}
  These are identical to the functions which determine the structure of the
approaching waves. The compatibility of these data at the point of collision
follows from a correct matching of the original travelling waves with the
Minkowski background. In this case, these data satisfy the conditions
  $$ {\cal E}_+(0)={\cal E}_-(0)=-1,\qquad \Phi_+(0)=\Phi_-(0)=0. $$
  Thus, these also provide normalization conditions for the Ernst potentials
at the point of collision: ${\cal E}(0,0)=-1$ and $\Phi(0,0)=0$.

\paragraph*{Geometrically defined null coordinates $(u,v)$ and $(\xi,\eta)$.}

In the above formulation of the characteristic initial value problem, it
seems appropriate to emphasize that two different pairs of geometrically
defined null coordinates can be used. The coordinates $(u,v)$ have been
adapted to the orbit space of the isometry group where the dynamics of waves
effectively takes place. These coordinates have been introduced in the
approaching wave interior regions, prior to their collision, as the affine
parameters on the plane wave rays and on the geodesics crossing the
corresponding wavefronts. In the wave interaction region, these coordinates
can then be defined uniquely using their continuity on the initial
characteristics. However, in this region, these coordinates lose their
property as affine parameters.

It is also convenient to introduce another system of null coordinates
$(\xi,\eta)$. These are also defined geometrically but, unlike the $(u,v)$
coordinates which are adapted to the orbit space of the isometry group, the
definition of these coordinates is adapted to the internal geometry of the
orbits themselves and, therefore, to the internal geometry of the
wave surfaces. More importantly for us, the coordinates $(\xi,\eta)$ are
better adapted to the analysis of the field equations.

The relation between these two coordinate systems can be found using the
definition (\ref{xietadef}) and the expression
(\ref{Alpha}) derived above. We can take the function $\beta$, which is
conjugate to $\alpha$, as
  $$ \beta=\alpha_+(u)-\alpha_-(v). $$
  A constant of integration in this expression has been chosen to give
$\beta=0$ at the point of collision. With this expression, $\xi$ and $\eta$
have the forms:
     \begin{equation}\label{xietauv}
     \xi(u)=2\alpha_+(u)-1, \qquad
     \eta(v)=1-2\alpha_-(v).
     \end{equation}
  Since $\alpha_+(u)$ and $\alpha_-(v)$ are monotonically decreasing
functions, $\xi$ and $\eta$ can be adopted as coordinates throughout the
interaction region, with $\xi=1$ and $\eta=-1$ on the wavefronts
$u=0$ and $v=0$ respectively.

Unfortunately, when formulated in the coordinates $(\xi,\eta)$, the
characteristic initial value problem under consideration possesses some
unpleasant properties. The problem arises from the fact that these
coordinates, when extended to the initial regions, take the forms $\xi=1$
for $u\le0$ and $\eta=-1$ for $v\le0$, so that the expressions for the
inverse coordinate transformations $u(\xi)$ and $v(\eta)$ have
nonanalytic behaviour near the wavefronts. This means that, even for
physically acceptable solutions which possess no singularities on the
wavefronts and which, therefore, are described by regular functions of
the coordinates $u$ and $v$, the metric components and potentials
possess some singular behaviour near the wavefronts when expressed
in terms of the coordinates $\xi$ and $\eta$. In particular, although they
are continuous on the wavefronts, they possess unbounded $\xi$- or
$\eta$-derivatives there.

\section{Monodromy transform and the integral evolution equations}

In our present construction, we use the general approach developed in
\cite{Alekseev:1985}, \cite{Alekseev:1987} and \cite{Alekseev:2000} for the
analysis of the structure and construction of solutions for integrable
hyperbolic reductions of the vacuum Einstein and electrovacuum
Einstein--Maxwell field equations. In this approach, which is called the
monodromy transform (because of its analogy with the well known inverse
scattering transform), every  solution of the reduced field equations is
characterized by two (for vacuum) or by four (for electrovacuum) functions
which depend on the spectral parameter only. These are interpreted as the
monodromy data of the corresponding fundamental solution of the associated
linear system. The problem of the solution of the nonlinear field equations
has been substituted in this approach by a solution of some linear singular
integral equations whose scalar kernel is built from these monodromy data.

A further development of this approach, which provides some new mathematical
tools for the solution of characteristic initial value problems, has been
presented in \cite{Alekseev:2001}. In that paper, the general solution of
the associated spectral problem was represented by two scattering matrices.
These dress the initial values of the fundamental solution of the associated
linear system given on two characteristics which pass through some
arbitrarily chosen point where the solutions are normalized. The monodromy
structure of these scattering matrices is characterised by ``dynamical''
monodromy data whose evolution is determined completely by the initial data
for the fields on these  characteristics. In this context, the problem of
the solution of the nonlinear field equations reduces to a solution of some
linear integral ``evolution'' equations of quasi-Fredholm type. The scalar
kernels of these equations are built from the dynamical monodromy data and
the  initial values of the fundamental solution of the associated linear
system on the above-mentioned initial characteristics. In terms of solutions
of these linear integral evolution equations, the corresponding solution of
the nonlinear field equations can be calculated in quadratures. Thus, many
properties which are revealed by these constructions are in principle well
adapted for the solution of the characteristic initial value problem for
various integrable reductions of Einstein's equations.

As formulated in the cited papers, however, these constructions cannot be
applied to a solution of the colliding plane wave problem. This is because
the local analyticity of the solution is violated on the wavefronts.
Moreover, as mentioned at the end of the previous section, although the
components of physically acceptable solutions are regular functions of the
$(u,v)$ coordinates, their $\xi$- and $\eta$-derivatives have singular
behaviour on the wavefronts. Fortunately, however, it has been found
possible to derive a generalization of the previous construction which
overcomes this difficulty. This was reported briefly in
\cite{Alekseev-Griffiths:2001} where the corresponding generalized linear
integral evolution equations were presented. It was shown how the character
of the singularities of the kernels of the quasi-Fredholm integral equations
can be explicitly related to that of the characteristic initial data. This
construction, whose detailed description and some applications we present
below, remains valid and, eventually, opens the way for the consideration of
a wide range of types of colliding waves with nonanalytical behaviour on
their wavefronts.

\subsection{Associated linear system for the Ernst equations}

Consider a linear system for a complex $3\times3$ matrix function
$\bigpsi(\xi,\eta,w)$ which depends on two real null coordinates $\xi$,
$\eta$ (as mentioned above) and a free complex (spectral) parameter $w$. To
represent the electrovacuum Ernst equations, this system should be
supplemented by additional conditions of the following two kinds. The first
consists of algebraic restrictions on the matrix coefficients of the system.
Together with the linear system itself, this can be presented in the form
     \begin{equation}
     \left\{\begin{array}{lclcl}
       \partial_\xi\bigpsi= \displaystyle{\frac {\U(\xi,\eta)}
       {2i(w-\xi)}}\bigpsi,&&
       \mbox{rank}\,\U=1, && \mbox{tr}\,\U=i,\\
       \partial_\eta\bigpsi=
\displaystyle{\frac{\V(\xi,\eta)}{2i(w-\eta)}}\bigpsi,&&
       \mbox{rank}\,\V=1, && \mbox{tr}\,\V=i \end{array} \right.
     \label{UVEqs}
     \end{equation}
The other conditions imply the existence of a Hermitian matrix integral of
(\ref{UVEqs}) which possesses the following structure
  \begin{equation} \left\{
\begin{array}{l} \bigpsi^\dagger\>\W\> \bigpsi = \W_0(w),\\
       \W_0^\dagger(w)=\W_0(w), \end{array} \qquad {\partial\W\over
        \partial w}=4 i{\bf\Omega},\right. \qquad
        {\bf\Omega}=\begin{pmatrix} 0&1&0\\ -1&0&0\\
0&0&0\end{pmatrix}\label{WEqs}
  \end{equation}
  where the Hermitian conjugation ${}^\dagger$ is defined as
$\W_0^\dagger(w)\equiv\overline{\W^T_0(\overline{w})}$, and $\W_0(w)$ is an
arbitrary (nondegenerate) Hermitian $3\times3$ matrix function of $w$.

\paragraph*{Equivalence of the matrix problem to the field equations.}

The conditions (\ref{UVEqs}), (\ref{WEqs}) are equivalent to the  symmetry reduced Einstein--Maxwell equations \cite{Alekseev:1987,Alekseev:2000}. For pure vacuum gravitational fields, the same set of conditions arise but for $2\times2$ matrices for which the third rows and columns of all matrices in (\ref{UVEqs}) and (\ref{WEqs}) are omitted. In the fully general case, these conditions imply that the matrices $\U$ and $\V$ have the specific forms
     \begin{equation}
     \label{UVMatrices}
     \begin{array}{lccl}     
\U={\cal F}_+\cdot\widehat{\U}\cdot{\cal F}_+^{-1}, \qquad
\widehat{\U}=\left(\begin{matrix} 1\\0\\0 \end{matrix}\right)
\otimes (i,\, -\partial_\xi{\cal E},\,\partial_\xi\Phi), \\[3ex]
\V={\cal F}_-\cdot\widehat{\V}\cdot{\cal F}_-^{-1}, \qquad
\widehat{\V}=\left(\begin{matrix} 1\\0\\0 \end{matrix}\right)
\otimes (i,\,-\partial_\eta{\cal E},\,\partial_\eta\Phi),
    \end{array}\qquad 
    {\cal F}_\pm =
\begin{pmatrix}
1&0&0\\
p_\pm&1&0\\
q_\pm&0&1 \end{pmatrix},
    \end{equation}
     in which $p_\pm=\Omega\pm \displaystyle\frac {i\alpha} H$ and
$q_\pm=2\overline{\widetilde\Phi}-2\overline{\Phi}
\left(\Omega\pm \displaystyle{i\alpha\over
H}\right)$. Also, the matrix $\W$ must be a linear function of the spectral parameter
$w$ with the structure
   \begin{equation}
   \W=4 i(w-\beta){\bf\Omega} +\G,\qquad
\G=\begin{pmatrix}\label{WMatrix}
4(H\Omega^2+\displaystyle\frac{\alpha^2} H)+4\widetilde{\Phi}
\overline{\widetilde{\Phi}}& -4 H \Omega-4\widetilde{\Phi}
\overline{\Phi}& -2 \widetilde{\Phi}\cr
-4 H \Omega-4\overline{\widetilde{\Phi}}\Phi& 4
H+4\Phi\overline{\Phi}&2\Phi\cr
-2\overline{\widetilde{\Phi}}&2\overline{\Phi}& 1
     \end{pmatrix}
     \end{equation}
  If in (\ref{UVMatrices})--(\ref{WMatrix}), $\alpha$, $\beta$, $H$,
$\Omega$, ${\cal E}$, $\Phi$ and $\widetilde{\Phi}$, are identified
respectively with the previously defined functions $\alpha\equiv \frac
12(\xi-\eta)$, $\beta\equiv\frac 12(\xi+\eta)$, the metric functions $H$ and
$\Omega$ in (\ref{OrbitMetric}), the Ernst potentials ${\cal E}$ and
$\Phi\equiv\Phi_y$, and another component $\widetilde{\Phi}\equiv\Phi_z$ of
the complex electromagnetic potential, then (\ref{UVEqs})--(\ref{WEqs})
imply that all the field equations and relations given above are satisfied
for these functions. The inverse statement is also true, so that all of the
functions in (\ref{UVMatrices})--(\ref{WMatrix}) can be calculated for any
solution of the reduced Einstein--Maxwell equations. Thus, the set of
conditions (\ref{UVEqs})--(\ref{WEqs})  for the four complex matrix functions
     $$ \bigpsi(\xi,\eta,w), \quad \U(\xi,\eta), \quad \V(\xi,\eta), \quad
\W(\xi,\eta,w) $$
  constitute a problem whose solution is equivalent to the solution of the
symmetry reduced Einstein--Maxwell equations. However, there is a
significant advantage in substituting an analysis of the matrix problem
(\ref{UVEqs})--(\ref{WEqs}) in place of studying the nonlinear partial
differential equations which govern the collision of gravitational or
gravitational and electromagnetic waves. It was just this step that was made
in  previous formulations of the monodromy transform approach, and we
develop this further here.

\subsection{Normalization conditions}

Before we analyze the matrix problem (\ref{UVEqs})--(\ref{WEqs}), it is
important to remove the gauge freedom which exists in these equations.
Without loss of  generality, we can impose normalization conditions for
the values at some ``initial'' or ``reference'' space-time point of the matrix
functions $\bigpsi$ and $\W$, as well as of the components of the metric
and electromagnetic potentials and the Ernst potentials. For the problem
considered here, it is convenient to choose this point of normalization as the point at which the waves collide. i.e. with the point $u=0$, $v=0$, or $\xi=1$, $\eta=-1$. First, we recall that we have already normalized the metric functions and the Ernst potentials at $u=0$, $v=0$ as follows:
  $$\begin{array}{lclclcl}
\alpha(0,0)=1,&& H(0,0)=1,&& {\cal E}(0,0)=-1, 
&&\widetilde{\Phi}(0,0)=0 \\[1ex]
\beta(0,0)=0, && \Omega(0,0)=0,&&\Phi(0,0)=0. &&\end{array} $$
For the matrix function $\bigpsi$, we can adopt the normalization
     \begin{equation}
     \label{PsiNorm}
  \bigpsi(1,-1,w)=\I.
     \end{equation}
 by using the transformation
$\bigpsi(\xi,\eta,w)\to\bigpsi(\xi,\eta,w)\bigpsi^{-1}(1,-1,w)$. 

 From (\ref{WEqs}) and (\ref{WMatrix}), it is now easy to show that the above
normalization conditions imply that the matrix
integral $\W_0(w)$ in (\ref{WEqs}) satisfies
     $$ \begin{array}{l}
\W_0(w)=4 iw{\bf\Omega}
+\mbox{diag}\,(4,4,1)\end{array}. $$

\subsection{The analytical structure of $\bigpsi$ on the spectral plane}

Everywhere below, $\bigpsi$ denotes the fundamental solution of the linear system (\ref{UVEqs}), normalized at the point of collision as in
(\ref{PsiNorm}). The general analytical structure of this solution was
described in  detail in \cite{Alekseev:1985}, \cite{Alekseev:1987},
\cite{Alekseev:2000}. However, in those formulations, $\bigpsi$ was
normalized at some arbitrarily chosen space-time point at which all
components of the solution of the field equations are analytic functions of the coordinates $\xi$ and $\eta$. However, in the present case we
specifically choose the point of normalization to be the point
$(\xi,\eta)=(1,-1)$ at which the waves collide. It is obvious that this
point is not a point at which the field components are locally analytic.

     $$ \begin{matrix}
\hskip1ex
L_{\scriptscriptstyle-} \hskip17ex L_{\scriptscriptstyle+} \cr
\noalign{\vskip-1.5ex}
{\vrule width10ex height0.05ex depth0.05ex}
{\vrule width0.05ex height0.5ex depth0.5ex}
{\vrule width10ex height0.15ex depth0.15ex}
{\vrule width0.05ex height0.5ex depth0.5ex}
{\vrule width10ex height0.05ex depth0.05ex}
{\vrule width0.05ex height0.5ex depth0.5ex}
{\vrule width10ex height0.15ex depth0.15ex}
{\vrule width0.05ex height0.5ex depth0.5ex}
{\vrule width10ex height0.05ex depth0.05ex}\cr
\noalign{\vskip0.1ex}
\hskip-1ex -1
\hskip8ex \eta \hskip10ex \xi \hskip8ex 1
  \end{matrix}$$
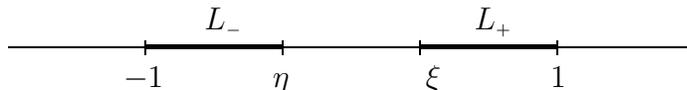
  $$\hbox{Figure 3:\quad \small Cuts in the spectral plane $w$}$$

As in the regular case, with the normalization (\ref{PsiNorm}), the structure
of the associated linear system implies that the fundamental solution
$\bigpsi(\xi,\eta,w)$ (and also its inverse) possesses four branch points on
the spectral plane $w$. The relative positions of these points and the cuts
$L_\pm$ which join them are indicated in Fig. 3. These singular points
include the two ``variable'' points $w=\xi$ and $w=\eta$ which correspond to
the poles of the coefficients of the linear system (\ref{UVEqs}). As in the
regular case, these singularities for $\bigpsi$ are branch points of order
$-1/2$. In addition, there are two fixed branch points of $\bigpsi$ at $w=1$
and $w=-1$ which arise from the normalization (\ref{PsiNorm}) imposed at
the point of the collision. However, the order of the branching of $\bigpsi$
at these two points may differ from
$1/2$ as occurs in the regular case. Moreover, the character of these branch
points is not universal in this case. In fact, it becomes dependent on the
solutions and differs from one solution to another.

Fortunately, we can proceed with our analysis without needing to specify the
order of branching of $\bigpsi$ at the points  $w=1$ and $w=-1$ on the spectral
plane. Following mainly the same simple arguments given in
\cite{Alekseev:1985,Alekseev:1987} and the ideas suggested for the regular
case in \cite{Alekseev:2001}, we present below a generalization of this
construction that is available for the analysis of fields which have
nonanalytic behaviour on the characteristics.

Because the rank of the matrix coefficients of the linear equations
(\ref{UVEqs}) is everywhere not more than $1$, near each of the singular
points of $\bigpsi$ only one of three linearly independent solutions of the
associated linear system can be branching, while the two others should be
holomorphic. This immediately implies that  the normalized fundamental
solution $\bigpsi$  in general has the local structure on the cuts $L_\pm$
which can be expressed in the forms:
  \begin{equation}
  \label{LocalA}
\begin{array}{ll}
L_+:& \bigpsi(\xi,\eta,w)=
\widetilde{\bpsi}_+(\xi,\eta,w) \otimes\k_+(w)
+\M_+(\xi,\eta,w)\\[2ex]
L_-:& \bigpsi(\xi,\eta,w)=
\widetilde{\bpsi}_-(\xi,\eta,w) \otimes\k_-(w)
+\M_-(\xi,\eta,w)
\end{array}
\end{equation}
  where, as in the regular case, the coordinate independent components of the
row vectors $\k_+(w)$ and $\k_-(w)$ are regular near the cuts $L_+$ and
$L_-$ respectively or, at least, are not branching. However, they can have
isolated singularities at the endpoints $w=1$ and $w=-1$ respectively. The
same is true for the matrices $\M_+$ and $\M_-$ on $L_+$ and $L_-$
respectively. In the regular case, the vectors
$\widetilde\bpsi_\pm(\xi,\eta,w)$ always have the structures
$\widetilde{\bpsi}_+(\xi,\eta,w) =\sqrt{w-1\over w-\xi}
\bpsi_+(\xi,\eta,w)$ and $\widetilde{\bpsi}_-(\xi,\eta,w)=\sqrt{w+1\over
w-\eta} \bpsi_-(\xi,\eta,w)$ where the vectors $\bpsi_+$ and 
$\bpsi_-$ are holomorphic on the cuts $L_+$ and $L_-$ respectively. 
In the general case, the components of the column vectors 
$\widetilde{\bpsi}_\pm(\xi,\eta,w)$ also have branch points at the 
endpoints of the corresponding cuts $L_\pm$. However, the characters 
of their singularities at the points $w=1$ and $w=-1$ respectively 
can be different for different solutions. Their specific characters 
are determined by the physical properties of the wavefronts as 
incorporated in the characteristic initial data for the approaching 
waves.

\subsection{Characteristic initial value problem for $\bigpsi$ and 
analytical structure of its initial data}

Following a similar construction as for the regular case given in
\cite{Alekseev:2001}, we now reformulate the characteristic initial 
value problem for the hyperbolic Ernst equations described in section 
\ref{CharIVP} as an equivalent characteristic initial value problem 
for the matrix function $\bigpsi$. Later it will be shown that the 
solution of this problem can be reduced to a solution of some simple 
linear integral equations.

First, it is easy to observe that the given initial data
(\ref{BoundaryData}) on the characteristics enable us to calculate the
matrix coefficients in (\ref{UVEqs}) on these characteristics. i.e. we can
determine the values of $\U(\xi,\eta=-1)$ and $\V(\xi=1,\eta)$. This enables
us to introduce two matrix functions $\bigpsi_+(\xi,w)$ and
$\bigpsi_-(\eta,w)$ which are the normalized fundamental solutions of the
associated linear system on the characteristics $\xi=1$ and $\eta=-1$:
     \begin{equation}\label{Psipm}
\left\{\begin{array}{l}
\partial_\xi\bigpsi_+(\xi,w)
=\displaystyle\frac{\U(\xi,-1)}{2i(w-\xi)}\cdot\bigpsi_+(\xi,w)\\[2ex]
\bigpsi_+(1,w)=\hbox{\bf I}\end{array}\right.\qquad
\left\{\begin{array}{l}
\partial_\eta\bigpsi_-(\eta,w)=
\displaystyle\frac{\V(1,\eta)}{2i(w-\eta)}
\cdot\bigpsi_-(\eta,w)\\[2ex]
\bigpsi_-(-1,w)=\hbox{\bf I}
     \end{array}\right.
     \end{equation}
     The coefficients of each of these systems are completely 
determined by the initial data for the fields on the corresponding 
characteristic. Thus these matrices form the characteristic initial 
data for the required solution for $\bigpsi(\xi,\eta,w)$:
     $$\bigpsi_+(\xi,w)\equiv\bigpsi(\xi,-1,w),
\quad \bigpsi_-(\eta,w)\equiv\bigpsi(1,\eta,w) \label{idata} $$

Applying now the same arguments that were used in \cite{Alekseev:1985,
Alekseev:1987}, we conclude that the boundary values $\bigpsi_+(\xi,w)$ and
$\bigpsi_-(\eta,w)$  should possess analytical structures on the spectral
plane which are very similar to that of $\bigpsi(\xi,\eta,w)$. Namely,
$\bigpsi_\pm(w=\infty)={\bf I}$ and $\bigpsi_+$ and its inverse are
holomorphic outside $L_+$, while $\bigpsi_-$ and its inverse are holomorphic
outside $L_-$. Moreover, their local structures on the cuts are given by the
expressions
     \begin{equation}
     \label{LocalB}
\begin{array}{lcl}
        L_+:&&\bigpsi_+(\xi,w)=
\widetilde{\bpsi}_{0+}(\xi,w)\otimes\k_+(w)+\M_{0+}(\xi,w),\\[2ex]
       L_-:&&\bigpsi_-(\eta,w)=
\widetilde{\bpsi}_{0-}(\eta,w)\otimes\k_-(w)+\M_{0-}(\eta,w)
        \end{array}
     \end{equation}
   where $\k_\pm(w)$ are the same as for $\bigpsi$ in (\ref{LocalA}), and
$\M_{0+}(\xi,w)$ and $\M_{0-}(\eta,w)$ are non-branching on $L_+$ and $L_-$
respectively. However, $\bigpsi_\pm$ may have isolated singularities at the
endpoints $w=1$ and $w=-1$ of these cuts respectively. As for the general
case described above, the vectors $\widetilde{\bpsi}_{0+}$ and
$\widetilde{\bpsi}_{0-}$ have branch points at the ends of the cuts $L_+$ or
$L_-$ respectively. Also, the character of their branching at these points
is completely determined by the initial data.

\subsection{The ``scattering'' matrices and their structures}

We now introduce, in analogy with the regular case,
the ``evolution'' or ``scattering'' matrices $\bchi_\pm(\xi,\eta,w)$, which represent $\bigpsi(\xi,\eta,w)$ in two alternative forms
       \begin{equation}
       \label{Dressing}
\bigpsi(\xi,\eta,w)=\bchi_+(\xi,\eta,w)\cdot
\bigpsi_+(\xi,w), \qquad
      \bigpsi(\xi,\eta,w)=\bchi_-(\xi,\eta,w)\cdot
\bigpsi_-(\eta,w).
      \end{equation}
   To understand the analytical structures of these matrices on the spectral
plane, we express them using the above relations in the form
\begin{equation}\label{Chiplusminus}
\bchi_+(\xi,\eta,w)\equiv\bigpsi(\xi,\eta,w)\cdot
\bigpsi_+^{-1}(\xi,w),\qquad
\bchi_-(\xi,\eta,w)\equiv\bigpsi(\xi,\eta,w)\cdot
\bigpsi_-^{-1}(\eta,w).
  \end{equation}
  It may be noted that these matrices are solutions of linear equations with
well defined initial conditions. We obtain these equations for $\bchi_\pm$
after multiplication of the second of two linear systems in (\ref{UVEqs}) by
$\bigpsi_+^{-1}(\xi,w)$ and the first one by $\bigpsi_-^{-1}(\eta,w)$ from
the right. We then find that each of the matrices $\bchi_+$ and $\bchi_-$
are determined by the linear systems with obvious initial conditions
     \begin{equation}
\left\{\begin{array}{l}
\partial_\xi\bchi_-
=\displaystyle\frac{\U(\xi,\eta)} {2i(w-\xi)}\cdot\bchi_-\\[2ex]
\bchi_-(1,\eta,w)=\hbox{\bf I}\end{array}\right.
\hskip1ex\left\{\begin{array}{l}
\partial_\eta\bchi_+
=\displaystyle\frac{\V(\xi,\eta)} {2i(w-\eta)}\cdot\bchi_+\\[2ex]
\bchi_+(\xi,-1,w)=\hbox{\bf I}\end{array}\right.
     \end{equation}
   For any given local solution of the Ernst equations each of these two
systems admits a unique solution. In both of these cases, the structures of the linear systems and the initial conditions show immediately that 
the corresponding fundamental solutions $\bchi_+$ or $\bchi_-$ should 
be holomorphic on the spectral plane $w$ everywhere outside the cut 
$L_-$ or $L_+$ respectively. Each of these fundamental solutions 
possesses only two branch points, which are located at the endpoints 
of the corresponding cut, and a finite jump on this cut. In addition, 
it is easy to show that these matrix functions should also possess 
the property $\bchi_\pm(\xi,\eta,w=\infty)\equiv\I$.

All of the above features enable us to represent each of these matrix
functions as the Cauchy integrals of their jumps over the corresponding cut:
   \begin{equation}
   \label{CauchyIntsA}
\bchi_+(\xi,\eta,w) =\I+\displaystyle\frac1{i\pi}\int\limits_{L_-}
\frac{[\bchi_+]_{\zeta_-}}{\zeta_--w}d\zeta_-\qquad
\bchi_-(\xi,\eta,w) =\I+\displaystyle\frac1{i\pi}\int\limits_{L_+}
\frac{[\bchi_-]_{\zeta_+}}{\zeta_+-w}d\zeta_+
   \end{equation}
   where $[\ldots]_\zeta$ is the jump (a half of the difference between the
left and right limits) of a function at the point $w=\zeta$ on a cut. It
may be noted that we usually choose directions on the paths $L_+$ and
$L_-$ shown in Fig. 3 from $w=1$ to $w=\xi$ and from $w=-1$ to $w=\eta$
respectively. However, the above definition of the jumps makes the
integrals (\ref{CauchyIntsA}) insensitive to the choice of these directions.
It is also necessary to note here that the convergence of the Cauchy
integrals in (\ref{CauchyIntsA}) at $w=-1$ and $w=1$ respectively has been
assumed. This conjecture has been proved for the regular case, and is
confirmed for all known examples which possess different degrees of
non-smooth behaviour on the characteristics.

Let us now explicitly calculate the structure of the jumps of $\bchi_\pm$
on the cuts using the definitions (\ref{Chiplusminus}), the local
structures (\ref{LocalA}) as well as the analyticity of $\bigpsi_+$ and its
inverse on $L_-$ and $\bigpsi_-$ and its inverse on $L_+$. Simple
calculations lead to the expressions
   \begin{equation}
   \label{JumpsA}
[\bchi_+]_{\tau_-}=[\widetilde{\bpsi}_-]_{\tau_-} \otimes
\m_-(\xi,\tau_-),\qquad
[\bchi_-]_{\tau_+}=[\widetilde{\bpsi}_+]_{\tau_+} \otimes
\m_+(\eta,\tau_-),
   \end{equation}
where $\tau_+\in L_+$ and $\tau_-\in L_-$. Substituting these into the integrands of (\ref{CauchyIntsA}), we obtain
   \begin{equation}
   \label{CauchyIntsB}
\begin{array}{l}\bchi_+(\xi,\eta,w)={\bf I} +\displaystyle{{1\over \pi
i}\int\limits_{L_-}{[\widetilde{\bpsi}_-]_{\zeta_-} \otimes
\m_-(\xi,\zeta_-)\over \zeta_--w}\, d\zeta_-}\\
\bchi_-(\xi,\eta,w)={\bf I}+\displaystyle{{1\over \pi
i}\int\limits_{L_+}{[\widetilde{\bpsi}_+]_{\zeta_+} \otimes
\m_+(\eta,\zeta_+)\over \zeta_+-w}\, d\zeta_+}.\end{array}
   \end{equation}
   These expressions involve a new, evolving kind of vector-functions
   \begin{equation}
     \label{DMData}
\m_+(\eta,w)=\k_+(w)\cdot\bigpsi_-^{-1}(\eta,w), \qquad
\m_-(\xi,w)=\k_-(w)\cdot\bigpsi_+^{-1}(\xi,w).
   \end{equation}
  These ``dynamical'' monodromy data replace the coordinate independent
vectors $\k_\pm(w)$ which arose earlier in similar integral
representations for $\bigpsi$ normalized at a regular point. (The
interpretation of these vector functions was explained in
\cite{Alekseev:2001} for the regular case.)

\subsection{Integral ``evolution'' equations}

The alternative representations of $\bigpsi$ that are given in
(\ref{Dressing}) implies an obvious consistency condition
   \begin{equation}
   \label{ConsistencyA}
\bchi_+(\xi,\eta,w)\bigpsi_+(\xi,w)= \bchi_-(\xi,\eta,w)\bigpsi_-(\eta,w).
   \end{equation}
   It is useful to recall that the matrix functions $\bigpsi_+$ and
$\bchi_-$ are holomorphic on the spectral plane everywhere outside $L_+$,
while $\bigpsi_-$ and $\bchi_+$ are holomorphic outside $L_-$. In addition, the values of these four matrices at $w=\infty$ are all equal to the unit matrix $\I$. Therefore, the right and left hand sides of
(\ref{ConsistencyA}) are analytical functions outside the cuts $L_+$ and
$L_-$. Hence, for the condition (\ref{ConsistencyA}) to be satisfied, it
is necessary that (here and everywhere below, $\tau_+\in L_+$ and $\tau_-\in L_-$)
   \begin{equation}
   \label{ConsistencyB}
\bchi_+(\tau_+)[\bigpsi_+]_{\tau_+}=
[\bchi_-]_{\tau_+}\bigpsi_-(\tau_+),\qquad
[\bchi_+]_{\tau_-}\bigpsi_+(\tau_-)=
\bchi_-(\tau_-)[\bigpsi_-]_{\tau_-}.
   \end{equation}
   Here (and in some expressions below) we omit for simplicity the parametric
dependence of the functions on the coordinates $\xi$ , $\eta$, or both $\xi$
and $\eta$.

We now substitute into (\ref{ConsistencyB}) the integral representations
(\ref{CauchyIntsB}), and the expressions (\ref{JumpsA}). The expressions for
the jumps of $\bigpsi_\pm$ on the cuts $L_\pm$ can then be derived easily
from the local representations (\ref{LocalB}):
   \begin{equation}\label{JumpsB}
[\bigpsi_+]_{\tau_+}=
[\widetilde{\bpsi}_{0+}]_{\tau_+}\otimes\k_+(\tau_+),
\quad\tau_+\in L_+\qquad
[\bigpsi_-]_{\tau_-}=
[\widetilde{\bpsi}_{0-}]_{\tau_-}\otimes\k_-(\tau_-).
   \end{equation}
   It is convenient to denote the jumps of the column-vector functions
$\widetilde{\bpsi}_\pm$ on the cuts $L_\pm$ as
   $$ \bfi_+(\xi,\eta,\tau_+)\equiv[\widetilde{\bpsi}_+]_{\tau_+},\qquad
\bfi_-(\xi,\eta,\tau_-)\equiv[\widetilde{\bpsi}_-]_{\tau_-}, $$
   and the initial values of these jumps as respectively
   $$ \bfi_{0+}(\xi,\tau_+)\equiv[\widetilde{\bpsi}_{0+}]_{\tau_+},
\qquad \bfi_{0-}(\eta,\tau_-)\equiv [\widetilde{\bpsi}_{0-}]_{\tau_-}. $$
   Substituting the expressions (\ref{CauchyIntsB}), (\ref{JumpsA}) and
(\ref{JumpsB}) into (\ref{ConsistencyB}), we arrive at the following
coupled pair of linear integral equations for the functions
$\bfi_+(\xi,\eta,\tau_+)$ and $\bfi_-(\xi,\eta,\tau_-)$, which are defined
on the cuts $L_+$ and $L_-$ respectively:
   \begin{equation}
   \left\{\begin{array}{l}\label{IntEqsA}
\bfi_+(\tau_+)- \displaystyle{\int\limits_{L_-}} S_+(\tau_+,\zeta_-)
\bfi_-(\zeta_-) \,d\zeta_- = \bfi_{0+}(\tau_+)\\
      \bfi_-(\tau_-)-\displaystyle{\int\limits_{L_+}}
S_-(\tau_-,\zeta_+) \bfi_+(\zeta_+) \,d\zeta_+ =
      \bfi_{0-}(\tau_-) \end{array}\right.
   \end{equation}
where $\tau_+,\zeta_+\in L_+$ and $\tau_-,\zeta_-\in L_-$. Also the 
scalar kernels of these equations take the forms
   \begin{equation}
   \label{KernelsA}
     S_+(\xi,\tau_+,\zeta_-) =\dfrac
{\Big(\m_-(\xi,\zeta_-)\cdot \bfi_{0+}(\xi,\tau_+)\Big)}{i\pi(\zeta_--\tau_+)}, \qquad
   S_-(\eta,\tau_-,\zeta_+) =\dfrac
{\Big(\m_+(\eta,\zeta_+)\cdot
\bfi_{0-}(\eta,\tau_-)\Big)}{i\pi(\zeta_+-\tau_-)}.
     \end{equation}

The equations (\ref{IntEqsA}) can easily be decoupled into two separate
equations for the vector functions $\bfi_+(\xi,\eta,\tau_+)$ and
$\bfi_-(\xi,\eta,\tau_-)$. Substituting $\bfi_-(\xi,\eta,\tau_-)$ from the
second equation in (\ref{IntEqsA}) into the first, and inversely, we obtain
the decoupled integral equations
   \begin{equation}
   \label{IntEqsB}
\begin{array}{l}\bfi_+(\tau_+)-\displaystyle{\int
\limits_{L_+}}{\cal F}_+(\tau_+,\zeta_+)\bfi_+(\zeta_+)
\,d\zeta_+=\f_+(\tau_+)\\[2ex]
\bfi_-(\tau_-)-\displaystyle{\int \limits_{L_-}}{\cal
F}_-(\tau_-,\zeta_-)\bfi_-(\zeta_-)\,d\zeta_-=\f_-(\tau_-)
   \end{array}
   \end{equation}
   in which the kernels and right hand sides are given by
   \begin{equation}
   \label{KernelsB}
   \begin{array}{l}
{\cal F}_+(\tau_+,\zeta_+)=\displaystyle{\int\limits_{L_-}} {\cal
S}_+(\tau_+,\chi_-){\cal S}_-(\chi_-,\zeta_+)\, d\chi_-,\\[2ex]
{\cal F}_-(\tau_-,\zeta_-)=\displaystyle{\int\limits_{L_+}} {\cal
S}_-(\tau_-,\chi_+){\cal S}_+(\chi_+,\zeta_-)\, d\chi_+,\\[2ex]
\f_+(\tau_+)=\bfi_{0+}(\tau_+)+\displaystyle{\int\limits_{L_-}}
{\cal S}_+(\tau_+,\chi_-)\bfi_{0-}(\chi_-)\,d\chi_-,\\[2ex]

\f_-(\tau_-)=\bfi_{0-}(\tau_-)+\displaystyle{\int\limits_{L_+}}
{\cal S}_-(\tau_-,\chi_+)\bfi_{0+}(\chi_+)\,d\chi_+.
\end{array}
   \end{equation}
However, although they are decoupled, these equations possess more
complicated structures than those in (\ref{IntEqsA}).

All the coefficients of the integral equations (\ref{IntEqsA}), and the
equivalent decoupled pair (\ref{IntEqsB}), are determined by the initial
data functions $\bigpsi_+(\xi,w)$ and $\bigpsi_-(\eta,w)$. These are
determined, in turn, by the characteristic initial data for the field
variables (e.g., the Ernst potentials). The relations of the coefficients
with the characteristic initial data will be clarified in more detail in
subsequent sections where we consider particular examples of the calculation procedure.

\section{Principal algorithm for solution of the problem}

\subsection{Specification of the initial data}

As explained in sections \ref{SingleWave} and \ref{CharIVP}, we may consider
the initial data to consist of the Ernst potentials, given as functions of
the affine parameters $u$ and $v$ along the initial null characteristics
  \begin{equation}
  \label{InDataA}
  \{{\cal E}_+(u),\Phi_+(u)\},\qquad \{{\cal
E}_-(v),\Phi_-(v)\}.
  \end{equation}
  Alternatively, the initial data may consist of the metric functions and the
complex electromagnetic potential (again as functions of the affine
parameters $u$ and $v$)
\begin{equation}
\label{InDataB} \{H_+(u),\Omega_+(u),\Phi_+(u)\}, \qquad
\{H_-(v),\Omega_-(v),\Phi_-(v)\}
\end{equation}
  In either case, we require the continuity conditions
\begin{equation}
\label{InDataC}
{\cal E}_\pm(0)=-1,\quad \Phi_\pm(0)=0,\quad
H_\pm(0)=1,\quad \Omega_\pm(0)=0.
\end{equation}
  Because of the choice of $u$ and $v$ as affine parameters on the initial
characteristics, for the conformal factor $f$, we have $f_+(u)=1$ and
$f_-(v)=1$. Also, the function $\alpha(u,v)$ is given explicitly by
(\ref{Alpha}), where the functions $\alpha_+(u)$ and $\alpha_-(v)$ are
determined from the linear equations with the well defined initial
conditions:
  \begin{equation}
  \label{EqnAlphaplus}
\left\{\begin{array}{l}
\alpha_+^{\prime\prime}-\alpha_+^\prime
\displaystyle{\frac {H_+^\prime}{H_+}}+ \displaystyle{\frac
{\alpha_+} {2 H_+}}\left[\displaystyle{\frac {\left\vert{\cal
E}_+^\prime+2\overline{\Phi}_+\Phi_+^\prime\right\vert^2}
{H_+}}+4\vert\Phi_+^\prime\vert^2\right]=0,\\[3ex]
\alpha_+(0)=1,\quad\alpha_+^\prime(0)=0.
\end{array}\right.
\end{equation}
where $H_+(u)\equiv -\mbox{Re}\,{\cal E}_+(u)
-\Phi_+(u)\overline{\Phi}_+(u)$ and
\begin{equation}
\label{EqnAlphaminus}
\left\{\begin{array}{l}
\alpha_-^{\prime\prime}-\alpha_-^\prime
\displaystyle{\frac {H_-^\prime}{H_-}}+ \displaystyle{\frac
{\alpha_-} {2 H_-}}\left[\displaystyle{\frac {\left\vert{\cal
E}_-^\prime+2\overline{\Phi}_-\Phi_-^\prime\right\vert^2}
{H_-^2}}+4 \vert\Phi_-^\prime\vert^2\right]=0,\\[3ex]
\alpha_-(0)=1,\quad\alpha_-^\prime(0)=0.
\end{array}\right.
\end{equation}
where $H_-(v)\equiv -\mbox{Re}\,{\cal E}_-(v)
-\Phi_-(v)\overline{\Phi}_-(v)$. 
Having obtained the functions $\alpha_+(u)$ and $\alpha_-(v)$, the functions
$\xi(u)$ and $\eta(v)$ must now be determined using the definitions
(\ref{xietadef}).

A nice feature of the above specification is that the functions
(\ref{InDataA}) or (\ref{InDataB}) can be taken as initial data functions
and chosen arbitrarily, provided only that the conditions (\ref{InDataC})
are satisfied. In this case, we have to solve the linear equations
(\ref{EqnAlphaplus}) and (\ref{EqnAlphaminus}) to find the corresponding
functions $\alpha_+(u)$ and $\alpha_-(v)$.

\subsection{Characteristic initial values for $\bigpsi$}

Everything is now ready for the calculation of the characteristic initial
values for $\bigpsi(u,v,w)$. From the initial data, we must calculate the
matrix coefficients $\U(\xi(u),-1)=\U(u,{v=0})$ and
$\V(1,\eta(v))=\V({u=0},v)$ of the linear systems (\ref{Psipm}), using the definitions (\ref{UVMatrices}) and expressing
them in terms of the $u,v$ coordinates. We should then solve the equations
(\ref{Psipm}) and determine the normalized
fundamental solutions $\bigpsi_+(u,w)$ and $\bigpsi_-(v,w)$.

It is clear, of course, that this step is crucial for a practical
realization of the method. In practice, however, for a great majority of the
initial data, it is found to be impossible to solve these equations
explicitly in a closed form. Nonetheless, it may be argued that there should
exist some infinite hierarchies of the initial data for which the solutions
of these linear systems can be found in a desired form. In this case, it may
be hoped that such cases correspond to sets of initial data which are
reasonably dense in the whole space of the initial data and at least contain
cases that are of particular physical significance.

\subsection{Construction of the integral evolution equations and their
solution}

Once the matrix functions $\bigpsi_+(u,w)$ and $\bigpsi_-(v,w)$ have been
found explicitly, we are able to calculate the fragments of their local
structures (\ref{LocalB}), i.e. the monodromy data vectors $\k_\pm(w)$,
the dynamical monodromy data vectors $\m_+(v,w)$ and
$\m_-(u,w)$ as well as the characteristic initial values
$\widetilde{\bpsi}_{0+}(u,w)$, $\widetilde{\bpsi}_{0-}(v,w)$
of the functions $\widetilde{\bpsi}_{+}(u,w)$,
$\widetilde{\bpsi}_{-}(v,w)$ and the initial values of their
jumps $\bfi_{0+}(u,w=\tau_+)$, $\bfi_{0-}(v,w=\tau_-)$ on
the cuts $L_+$ and $L_-$ respectively. This is all we
need to construct the kernels (\ref{KernelsA}) and the right
hand sides of the main integral evolution equations
(\ref{IntEqsA}) or the kernels and right hand sides of the
decoupled form (\ref{IntEqsB}) of these equations.

It is then necessary to solve the linear integral equations (\ref{IntEqsA})
or (\ref{IntEqsB}) with the corresponding kernels and right hand sides. This
is the second crucial step in our algorithm because, even in the case in
which we are able to find $\bigpsi_+(u,w)$ and $\bigpsi_+(u,w)$ explicitly,
there are no guarantees that the solution of the linear integral equations
(\ref{IntEqsA}) or (\ref{IntEqsB}) can be found in an explicit form.
However, if the particular forms of the coefficients of these integral
equations (derived for the specified initial data) enables us to find their
solution explicitly, then the Ernst potentials and all the components of the
metric and the electromagnetic vector potential in the wave interaction
region can be expressed in quadratures (or even explicitly) in terms of
these solutions.

\subsection{Calculation of the field components}

As in the regular case \cite{Alekseev:2001}, all components of the solution
of the characteristic initial value problem can be expressed in terms of
the matrices $\R_\pm(u,v)$ defined by the asymptotic expansions of the
scattering matrices
  \begin{equation}
  \label{RMatrices}
\bchi_\pm(u,v,w)=\mbox{\bf
I}+\frac 1 w\R_\pm(u,v)+O(\frac 1 {w^2}).
  \end{equation}
  The integral representations (\ref{CauchyIntsB}) suggest the following
expressions for the matrices $\R_\pm(u,v)$ in terms of solutions of the
integral evolution equations (\ref{IntEqsA}) or (\ref{IntEqsB}) and the
dynamical monodromy data vectors $\m_+(v,\tau_+)$ and $\m_-(u,\tau_-)$
  \begin{equation}
  \label{RExpressions}
  \begin{array}{l}
\R_+(u,v)=-\displaystyle{1\over \pi i}\int\limits_{L_-}\,
\bfi_-(u,v,\zeta_-)\otimes\m_-(u,\zeta_-)\,d\,\zeta_-\\[2ex]
\R_-(u,v)=-\displaystyle{1\over \pi i}\int\limits_{L_+}\,
\bfi_+(u,v,\zeta_+)\otimes\m_+(v,\zeta_+)\,d\,\zeta_+
\end{array}
  \end{equation}
  Using these expansions in (\ref{UVEqs}) and (\ref{WEqs}) leads to the
expressions for the matrices $\U$ and $\V$:
  \begin{equation}
  \label{UVValues}
\U(u,v)=\U(u,0)+2 i\partial_\xi\R_+,\qquad
\V(u,v)=\V(0,v)+2 i\partial_\eta\R_-,
  \end{equation}
  where we have to transform the $\xi$- and $\eta$-derivatives into $u$-  and
$v$-derivatives respectively. In this way we also obtain the alternative
forms of the matrix $\W$
  \begin{equation}
\label{WValue}
\W(u,v,w)=\W(u,0,w)-4 i (\bigomega\R_++\R_+^\dagger\bigomega)
=\W(0,v,w)-4 i(\bigomega\R_-+\R_-^\dagger\bigomega)
  \end{equation}
  where the constant matrix $\bigomega$ was defined in (\ref{WEqs}). 
In accordance with (\ref{WMatrix}), the values of the matrices $\W(u,0,w)$ and $\W(0,v,w)$ in (\ref{WValue}) are determined completely by the characteristic initial data for the fields.

The expressions (\ref{UVValues}) provide us with the following alternative
forms of the expressions for the Ernst potentials:
  \begin{equation}
\label{ErnstPotentialsA}
\begin{array}{l}
{\cal E}(u,v)={\cal E}(u,0)-2 i (\e_1\cdot\R_+(u,v)\cdot\e_2)={\cal 
E}(0,v)-2 i (\e_1\cdot\R_-(u,v)\cdot\e_2)\\[2ex]
\Phi(u,v)=\Phi(u,0)+2 i (\e_1\cdot\R_+(u,v)\cdot\e_3)=\Phi(0,v)+2 i 
(\e_1\cdot\R_-(u,v)\cdot\e_3)
\end{array}
\end{equation}
where $\e_1=\{1,0,0\}$, $\e_2=\{0,1,0\}$ and $\e_3=\{0,0,1\}$.
It may be helpful here to mention also some additional interesting relations
for the matrices $\R_\pm(u,v)$, such as
\begin{equation}
\begin{array}{lcccl}
{\cal E}(u,0)=-1-2 i (\e_1\cdot\R_-(u,0)\cdot\e_2),&&&&
\Phi(u,0)=2 i (\e_1\cdot\R_-(u,0)\cdot\e_3),\\[2ex]
{\cal E}(0,v)=-1-2 i (\e_1\cdot\R_+(0,v)\cdot\e_2),&&&&
\Phi(0,v)=2 i (\e_1\cdot\R_+(0,v)\cdot\e_3).
\end{array}
\end{equation}

For later reference, we give here also the expressions
(\ref{ErnstPotentialsA}) in a more explicit form
  \begin{equation}
\label{ErnstPotentialsB}
\begin{array}{l}
{\cal E}(u,v)=
{\cal E}_+(u)+\displaystyle{2\over\pi}\displaystyle{ \int\limits_{L_-}}
(\e_1\cdot\bfi_-(u,v,\zeta_-))(\m_-(u,\zeta_-)\cdot\e_2)
d\zeta_- \\
\phantom{{\cal E}(u,v)}={\cal E}_-(v)
+\displaystyle{2\over\pi}\displaystyle{ \int\limits_{L_+}}
(\e_1\cdot\bfi_+(u,v,\zeta_+))(\m_+(v,\zeta_+)\cdot\e_2)
d\zeta_+ \\[2ex]
\Phi(u,v)=\Phi_+(u)-\displaystyle{2\over\pi}\displaystyle{ \int\limits_{L_-}}
(\e_1\cdot\bfi_-(u,v,\zeta_-))(\m_-(u,\zeta_-)\cdot\e_3)
d\zeta_- \\
\phantom{\Phi(u,v)}=\Phi_-(v)-\displaystyle{2\over
\pi}\displaystyle{ \int\limits_{L_+}}
(\e_1\cdot\bfi_+(u,v,\zeta_+))(\m_+(v,\zeta_+)\cdot\e_3)
d\zeta_+
\end{array}
\end{equation}
where we have returned to our basic notations, with ${\cal E}_+(u)$,
$\Phi_+(u)$ and ${\cal E}_-(v)$, $\Phi_-(v)$ denoting the characteristic
initial data for the Ernst potentials on the characteristics.
A more detailed description of the various steps of this 
algorithm is given in the following section, where we present various 
particular examples.

\section{Examples of solutions}

\subsection{Collision of two gravitational impulses}

As a first example, let us consider the collision of two impulsive
gravitational waves. This case is well known. It is described by the
solution of Khan and Penrose \cite{Khan-Penrose:1971} if the polarizations of the approaching waves are aligned. Its generalization to the case when the waves possess nonaligned polarizations was found later by Nutku and Halil \cite{Nutku-Halil:1977}. We consider these cases here to illustrate that our algorithm works in this situation without introducing any essential difference between the linear (aligned) or nonlinear (nonaligned) cases.

\paragraph*{Gravitational impulses before their collision.}

The gravitational waves under consideration have impulsive components located prior to their collision on two characteristics, and the space-time behind these is again part of a Minkowski space-time. Although the metric functions are continuous on the wavefronts, some of their derivatives in directions which cross the wavefronts are discontinuous and the curvature possesses a distributional character. In this case, 
the equation (\ref{EqnAlphaplus}) for the left wave as well as the equation (\ref{EqnAlphaminus}) for the right wave should be solved together with the conditions that all projections (\ref{NPscalars}) of the Weyl tensor should vanish behind the wavefront for each of the approaching waves. The solutions  subject to the conditions (\ref{InDataC}) take the forms
  \begin{equation}
  \label{NHLwave}
  \begin{array}{lcccl}
\alpha_{\scriptscriptstyle +}(u)=
1-k_{\scriptscriptstyle +}^2 u^2&&&&\alpha_{\scriptscriptstyle -}(v)=
1-k_{\scriptscriptstyle -}^2 v^2,\\[1ex]
{\cal E}_{\scriptscriptstyle +}(u)=
-1-2k_{\scriptscriptstyle +} e^{-i\delta_{\scriptscriptstyle +}} u
-k_{\scriptscriptstyle +}^2 u^2,&&&&
{\cal E}_{\scriptscriptstyle -}(v)=
-1-2k_{\scriptscriptstyle -} e^{-i\delta_{\scriptscriptstyle -}} v
-k_{\scriptscriptstyle -}^2 v^2,\\[1ex]
\Phi_{\scriptscriptstyle +}(u)=0,\qquad
f_{\scriptscriptstyle +}(u)=1,&&&&
\Phi_{\scriptscriptstyle -}(v)=0,\qquad f_{\scriptscriptstyle -}(v)=1,
\end{array}
  \end{equation}
where $k_{\scriptscriptstyle\pm}$ and $\delta_{\scriptscriptstyle\pm}$ are arbitrary real constants. The corresponding expressions for
$H_{\scriptscriptstyle\pm}(u)$ and $\Omega_{\scriptscriptstyle\pm}(u)$ are
  \begin{equation}
  \label{NHLwaveA}
  \begin{array}{ll}
H_{\scriptscriptstyle +}(u)=
1+2u k_{\scriptscriptstyle +}\cos\delta_{\scriptscriptstyle +}
+k_{\scriptscriptstyle +}^2 u^2, \qquad
  &\Omega_{\scriptscriptstyle +}(u) =\displaystyle\frac
{2uk_{\scriptscriptstyle +}\sin\delta_{\scriptscriptstyle+}}
{H_{\scriptscriptstyle +}(u)} \\[12pt]
  H_{\scriptscriptstyle-}(v)=
1+2v k_{\scriptscriptstyle-}\cos\delta_{\scriptscriptstyle-}
+k_{\scriptscriptstyle -}^2 v^2,
  &\Omega_{\scriptscriptstyle-}(v) =-\displaystyle\frac
{2vk_{\scriptscriptstyle -}\sin\delta_{\scriptscriptstyle-}}
{H_{\scriptscriptstyle-}(v)}
  \end{array}
  \end{equation}
  It may be noted that the constants $k_{\scriptscriptstyle\pm}$ play the
role of amplitudes for these waves. They determine the scales (distances) in
the affine parameters $u$ and $v$ at which the waves can focus initially
parallel null geodesic rays which cross them. The parameters
$\delta_{\scriptscriptstyle\pm}$ determine the polarization of the waves. It
is clear that either $\delta_{\scriptscriptstyle+}$ or
$\delta_{\scriptscriptstyle-}$ can be transformed to zero by a rotation of
the ignorable coordinates $y$ and $z$. However, it is the difference
$\delta_{\scriptscriptstyle+} -\delta_{\scriptscriptstyle-} (\mbox{mod}2\pi)$
which has physical significance, determining the relative polarizations of
the approaching waves. However, we keep below both the parameters
$\delta_\pm$ to maintain a symmetry of the subsequent expressions.

Using (\ref{Alpha}) and (\ref{NHLwave}), we obtain the functions $\alpha$ and
$\beta$ in the interaction region as
  $$ \alpha(u,v)=1-k_{\scriptscriptstyle +}^2 u^2
-k_{\scriptscriptstyle -}^2 v^2, \qquad
\beta(u,v)=k_{\scriptscriptstyle -}^2
v^2-k_{\scriptscriptstyle +}^2 u^2. $$
  From this we obtain the relation between the null coordinates $\xi$ and
$\eta$ and $u$ and $v$ respectively:
   \begin{equation}
  \label{NullCoords}
\xi=1-2 k_{\scriptscriptstyle +}^2 u^2, \qquad
\eta=-1+2 k_{\scriptscriptstyle -}^2 v^2.
   \end{equation}

\paragraph*{Calculation of the ``in-states''
$\bigpsi_{\scriptscriptstyle+}(u,w)$ and
$\bigpsi_{\scriptscriptstyle-}(v,w)$.}

With the characteristic initial data (\ref{NHLwave}) together with (\ref{NHLwaveA}), we can calculate the matrix coefficients $\U(\xi,\eta=-1)$ and $\V(\xi=1,\eta)$ of the linear systems (\ref{Psipm}) on the characteristics $\eta=-1$ (i.e. $v=0$) and $\xi=1$ (i.e. $u=0$) respectively:
   \begin{equation}
  \label{UinVin}
   \begin{array}{l}
   \U(u,0)= -\displaystyle{\frac 1{2 i k_{\scriptscriptstyle +} u}}
   \begin{pmatrix}
   e^{-i\delta_{\scriptscriptstyle +}}+k_{\scriptscriptstyle +} u\\
   i(e^{-i\delta_{\scriptscriptstyle +}}-k_{\scriptscriptstyle +} u)\\
   0\end{pmatrix}\otimes \Bigl(1,\, i,\,0\Bigr)\\[2ex]
   \V(0,v)= -\displaystyle{\frac 1{2 i k_{\scriptscriptstyle -} v}}
   \begin{pmatrix}
   e^{-i\delta_-}+k_{\scriptscriptstyle -} v \\
   -i(e^{-i\delta_-}-k_{\scriptscriptstyle -} v)\\
   0\end{pmatrix}\otimes \Bigl(1,\,-i,\,0\Bigr)
   \end{array}
   \end{equation}
  Substituting differentiation with respect to $u$ and $v$ rather than with
respect to $\xi$ and $\eta$ in (\ref{Psipm}) and
using the coefficients (\ref{UinVin}), these equations can be solved
explicitly. The normalized fundamental solutions take the forms:
  \begin{equation}
  \label{NHPsi}
\begin{array}{l}
  \bigpsi_{+}=\displaystyle{\frac {\lambda_{\scriptscriptstyle +}^{-1}}
{2(w-1)}}
  \begin{pmatrix} w-1-2 k_{\scriptscriptstyle +}
e^{-i\delta_{\scriptscriptstyle +}} u\\ -i (w-1+2 k_{\scriptscriptstyle +}
e^{-i\delta_{\scriptscriptstyle +}}u)\\0\end{pmatrix} \otimes
\Bigl(1,\,i,\,0 \Bigr)+
\begin{pmatrix} \displaystyle\frac 1 2& -\displaystyle\frac i 2&0\\[2ex]
\displaystyle\frac i 2  &\displaystyle\frac 1 2&0\\[2ex]
0&0&1\end{pmatrix}
\\[6ex]
\bigpsi_{-}=\displaystyle{\frac {\lambda_-^{-1}}
{2(w+1)}} \begin{pmatrix} w+1+2 k_{\scriptscriptstyle-} e^{-i\delta_-}v\\
i(w+1-2 k_{\scriptscriptstyle-} e^{-i\delta_-}v)\\0\end{pmatrix} \otimes
\Bigl(1,\,-i,\,0\Bigr)+
\begin{pmatrix} \displaystyle\frac 1 2& \displaystyle\frac i 2&0\\[2ex]
-\displaystyle\frac i 2  &\displaystyle\frac 1 2&0\\[2ex]
0&0&1\end{pmatrix}
\end{array}
\end{equation}
while the inverse matrices are
\begin{equation}\label{NHPsim1}
\begin{array}{l}
\bigpsi_{+}^{-1}=\displaystyle{\frac 12}\lambda_{\scriptscriptstyle+}
\begin{pmatrix}1\\-i\\0
\end{pmatrix}
\otimes
\begin{pmatrix} 1 & i&0
\end{pmatrix}+\begin{pmatrix} \displaystyle{\frac{w-1+2
k_{\scriptscriptstyle+} e^{-i\delta_{\scriptscriptstyle +}}u}{2(w-1)}}
& -\displaystyle{\frac{i(w-1-2 k_{\scriptscriptstyle +}
e^{-i\delta_{\scriptscriptstyle +}} u)} {2 (w-1)}}&0 \\
  \displaystyle{\frac{i (w-1+2 k_{\scriptscriptstyle +}
e^{-i\delta_{\scriptscriptstyle +}} u)}{ 2(w-1)}}&
  \displaystyle{\frac{w-1-2 k_{\scriptscriptstyle +}
e^{-i\delta_{\scriptscriptstyle +}}u}{2(w-1)}}&0\\[2ex] 0&0&1
\end{pmatrix}
\\[6ex]
\bigpsi_-^{-1}=\displaystyle{\frac 12}\lambda_-
\begin{pmatrix}
1\\ i\\ 0
\end{pmatrix}
\otimes
\begin{pmatrix}
1 & -i&0
\end{pmatrix}
+\begin{pmatrix} \displaystyle{\frac {w+1- 2 k_{\scriptscriptstyle -}
e^{-i\delta_-}v}{2(w+1)}}&
  \displaystyle{\frac{i(w+1+2 k_{\scriptscriptstyle -}
e^{-i\delta_-}v)}{2(w+1)}}&0\\ -\displaystyle{\frac {i(w+1- 2
k_{\scriptscriptstyle -} e^{-i\delta_-} v)}{2(w+1)}}&
  \displaystyle{\frac{(w+1+2 k_{\scriptscriptstyle -}
e^{-i\delta_-}v)}{2(w+1)}}&0\\[2ex] 0&0&1
\end{pmatrix}
\end{array}
   \end{equation}
where the functions $\lambda_{\scriptscriptstyle +}(u,w)$ and
$\lambda_-(v,w)$ are determined by the expressions (see also
(\ref{NullCoords}))
   \begin{equation}
   \label{lambdadef}
  \lambda_{\scriptscriptstyle +}(u,w)=\sqrt{\displaystyle{\frac
{w-\xi(u)}{w-1}}}, \quad
\lambda_-(v,w)=\sqrt{\displaystyle{\frac {w-\eta(v)}{w+1}}}
   \end{equation}
  which satisfy the additional conditions
$\lambda_{\scriptscriptstyle+}(u,w=\infty)=1$ and
$\lambda_-(v,w=\infty)=1$.

\paragraph*{Construction of the integral evolution equations.}

For our construction, we now need to determine certain fragments of the
structures of these matrix functions on the cuts $L_\pm$ defined in
(\ref{LocalB}). These are the conserved monodromy data vectors
   $$ \k_{\scriptscriptstyle +}(w)=\begin{pmatrix}1,& i,\,0\end{pmatrix},\qquad
   \k_-(w)=\begin{pmatrix} 1,& -i,\,0\end{pmatrix} $$
   and the initial values for the vector functions whose jumps on the 
cuts $L_{\scriptscriptstyle +}$ and $L_-$ respectively are the 
unknowns in
the integral evolution equations:
   $$ \begin{array}{l}
\widetilde{\bpsi}_{0+}(u,w)=\displaystyle{\frac 
{\lambda_{\scriptscriptstyle +}^{-1}}
{2(w-1)}} \begin{pmatrix} w-1-2k_{\scriptscriptstyle +} 
e^{-i\delta_{\scriptscriptstyle +}} u\\
-i (w-1+2k_{\scriptscriptstyle +} e^{-i\delta_{\scriptscriptstyle 
+}}u)\\0\end{pmatrix}\\[4ex]
\widetilde{\bpsi}_{0-}(v,w)=\displaystyle{\frac{\lambda_-^{-1}}{
2(w+1)}}\begin{pmatrix} w+1+2 k_{\scriptscriptstyle -} e^{-i\delta_-}v\\
i(w+1-2 k_{\scriptscriptstyle -} e^{-i\delta_-}v)\\0\end{pmatrix}
\end{array} $$

We now calculate the dynamical monodromy data vectors using the definitions
(\ref{DMData}), giving
   $$ \begin{array}{l}
\m_{\scriptscriptstyle +}(v,\zeta_{\scriptscriptstyle +})=\begin{pmatrix}
1-\displaystyle{\frac{ 2 k_{\scriptscriptstyle -}
e^{-i\delta_-} v}{\zeta_{\scriptscriptstyle +}+1}},&i\bigl(1+
\displaystyle{\frac{2 k_{\scriptscriptstyle -}
e^{-i\delta_-} v}{\zeta_{\scriptscriptstyle +}+1}}\bigr),& 
0\end{pmatrix}\\[2ex]
\m_-(u,\zeta_-)=\begin{pmatrix} 1 +\displaystyle{\frac{2 
k_{\scriptscriptstyle +}
e^{-i\delta_{\scriptscriptstyle +}} u}{\zeta_--1}},& -i
\bigl(1-\displaystyle{\frac{2 k_{\scriptscriptstyle +} 
e^{-i\delta_{\scriptscriptstyle +}}
u}{\zeta_--1}}\bigr),&0\end{pmatrix}\end{array} $$
   and the kernels $S_\pm$ defined in (\ref{KernelsA}), for which we 
obtain the expressions
   $$
S_{\scriptscriptstyle +}(u,v,\tau_{\scriptscriptstyle 
+},\zeta_-)=-\displaystyle{\frac 1{i\pi}}
\displaystyle {\frac{2 k_{\scriptscriptstyle +} 
e^{-i\delta_{\scriptscriptstyle +}} u
[\lambda_{\scriptscriptstyle +}^{-1}]_{\tau_{\scriptscriptstyle +}}}
{(\zeta_--1)(\tau_{\scriptscriptstyle +}-1)}},\qquad
S_-(u,v,\tau_-,\zeta_{\scriptscriptstyle +})=\displaystyle{\frac 1{i\pi}}
\displaystyle{\frac{2 k_{\scriptscriptstyle -} e^{-i\delta_-} v
[\lambda_-^{-1}]_{\tau_-}} {(\zeta_{\scriptscriptstyle +}+1)(\tau_-+1)}}.
 $$
   For this case, the kernels ${\cal F}_\pm$ and the right hand sides
$\f_\pm$  of the decoupled integral evolution equations (\ref{IntEqsB})
defined in (\ref{KernelsB}) take the forms
   $$ \begin{array}{l}
{\cal F}_{\scriptscriptstyle +}(u,v,\tau_{\scriptscriptstyle 
+},\zeta_{\scriptscriptstyle +})=-\displaystyle{\frac 1{
i\pi}} \displaystyle{\frac{2 k_{\scriptscriptstyle 
+}k_{\scriptscriptstyle -} e^{i(-\delta_{\scriptscriptstyle 
+}-\delta_-)}
u v }{\widetilde{v}
(\tau_{\scriptscriptstyle +}-1)(\zeta_{\scriptscriptstyle +}+1)}}\,[\lambda_{\scriptscriptstyle +}^{-1}]_{\tau_{\scriptscriptstyle 
+}}\\[3ex]
{\cal F}_-(u,v,\tau_-,\zeta_-)=\displaystyle{\frac 1{i\pi}}
\displaystyle{\frac{2k_{\scriptscriptstyle +}k_{\scriptscriptstyle -} 
e^{i(-\delta_{\scriptscriptstyle +}-\delta_-)} u v
}
{\widetilde{u} (\tau_-+1)(\zeta_--1)}}\,[\lambda_-^{-1}]_{\tau_-}\\[3ex]
\f_{\scriptscriptstyle +}(\tau_{\scriptscriptstyle +})=
{\frac 1 2}[\lambda_{\scriptscriptstyle 
+}^{-1}]_{\tau_{\scriptscriptstyle +}}\left[\begin{pmatrix}1\\
-i\end{pmatrix}-\displaystyle\frac{ 2k_{\scriptscriptstyle +} 
e^{-i\delta_{\scriptscriptstyle +}}
u}{\widetilde{v} (\tau_{\scriptscriptstyle +}-1)}
\begin{pmatrix}1+k_{\scriptscriptstyle -} e^{-i\delta_-} v\\ i
(1-k_{\scriptscriptstyle -} e^{-i\delta_-} v)\end{pmatrix}\right]\\[3ex]
\f_-(\tau_-)=
\frac 1 2[\lambda_-^{-1}]_{\tau_-}\left[\begin{pmatrix}1\\
i\end{pmatrix}-\displaystyle\frac{ 2k_{\scriptscriptstyle -} e^{-i\delta_-}
v}{\widetilde{u} (\tau_-+1)}
\begin{pmatrix}-1-k_{\scriptscriptstyle +} 
e^{-i\delta_{\scriptscriptstyle +}} u\\ i
(1-k_{\scriptscriptstyle +} e^{-i\delta_{\scriptscriptstyle +}} 
u)\end{pmatrix}\right]
   \end{array} $$
   where $\widetilde{u}=\sqrt{1-k_{\scriptscriptstyle +}^2 u^2}$ and
$\widetilde{v}=\sqrt{1-k_{\scriptscriptstyle -}^2 v^2}$.

\paragraph*{Solution of the integral evolution equations.}

The equations (\ref{IntEqsA}) and (\ref{IntEqsB}) show unambiguously that the
unknown functions $\bfi_\pm$ should possess the following structure:
   $$
\bfi_{\scriptscriptstyle +}(\tau_{\scriptscriptstyle 
+})=[\lambda_{\scriptscriptstyle +}^{-1}]_{\tau_{\scriptscriptstyle 
+}}\left\{\displaystyle\frac 1 2
\begin{pmatrix}1\\
-i\end{pmatrix}+\displaystyle\frac 1{\tau_{\scriptscriptstyle +}-1}
\begin{pmatrix}A_{\scriptscriptstyle +}\\ B_{\scriptscriptstyle 
+}\end{pmatrix}\right\},\quad
\bfi_-(\tau_-)=[\lambda_-^{-1}]_{\tau_-}\left\{\displaystyle\frac 1 2
\begin{pmatrix}1\\
i\end{pmatrix}+\displaystyle\frac 1{\tau_-+1}
\begin{pmatrix} A_-\\ B_-\end{pmatrix}\right\},
$$
   where $A_\pm$ and $B_\pm$ are unknown functions of $u$ and $v$.
Substituting these expressions into (\ref{IntEqsA}) and explicitly
calculating the integrals using analytical continuations and the elementary
theory of residues leads to algebraic equations whose solution is
   $$ \begin{array}{lcccl}
A_{\scriptscriptstyle +}=-\displaystyle\frac {k_{\scriptscriptstyle +}
e^{-i\delta_{\scriptscriptstyle +}}u(\widetilde{u}+k_{\scriptscriptstyle -}
e^{-i\delta_-} v)}
{\widetilde{u}\widetilde{v}-k_{\scriptscriptstyle 
+}k_{\scriptscriptstyle -} e^{i(-\delta_{\scriptscriptstyle 
+}-\delta_-)} u v} &&&&
A_-=\displaystyle\frac{k_{\scriptscriptstyle -} 
e^{-i\delta_-}v(\widetilde{v}+k_{\scriptscriptstyle +}
e^{-i\delta_{\scriptscriptstyle +}} u)}
{\widetilde{u}\widetilde{v}-k_{\scriptscriptstyle 
+}k_{\scriptscriptstyle -} e^{i(-\delta_{\scriptscriptstyle 
+}-\delta_-)} u v}\\[2ex]
B_{\scriptscriptstyle +}=-\displaystyle\frac {i k_{\scriptscriptstyle 
+} e^{-i\delta_{\scriptscriptstyle +}}
u(\widetilde{u}-k_{\scriptscriptstyle -} e^{-i\delta_-} v)}
{\widetilde{u}\widetilde{v}-k_{\scriptscriptstyle 
+}k_{\scriptscriptstyle -} e^{i(-\delta_{\scriptscriptstyle 
+}-\delta_-)} u v}&&&&
B_-=-\displaystyle\frac{i k_{\scriptscriptstyle -}
e^{-i\delta_-}v(\widetilde{v}-k_{\scriptscriptstyle +} 
e^{-i\delta_{\scriptscriptstyle +}} u)}
{\widetilde{u}\widetilde{v}-k_{\scriptscriptstyle 
+}k_{\scriptscriptstyle -} e^{i(-\delta_{\scriptscriptstyle 
+}-\delta_-)}
u v}
\end{array} $$

\paragraph*{Calculation of the Ernst potential.}

   It remains now to substitute this solution into the general expression
(\ref{ErnstPotentialsB}) for the Ernst potential ${\cal E}$. Calculating the
corresponding integral leads to the expression
   $$ {\cal E}(u,v)=1-k_{\scriptscriptstyle +}^2 u^2
-k_{\scriptscriptstyle-}^2 v^2
-\displaystyle{\frac{2(\widetilde{u}+k_{\scriptscriptstyle-} v
e^{-i\delta_-})(\widetilde{v}+k_{\scriptscriptstyle +} u
e^{-i\delta_{\scriptscriptstyle +}})}{
\widetilde{u}\widetilde{v} -k_{\scriptscriptstyle+}k_{\scriptscriptstyle-}
e^{i(-\delta_{\scriptscriptstyle +}-\delta_-)} u v}} $$
   It is not difficult to check directly that this solution satisfies all of
the characteristic initial conditions of the problem under consideration
and that it coincides with the corresponding expression for the Nutku--Halil
solution. The case $\delta_{\scriptscriptstyle+}
=\delta_{\scriptscriptstyle-}=0$ can be seen to correspond to the Ernst
potential of the Khan--Penrose solution.

\subsection{Collision of gravitational impulse with electromagnetic
step-like wave of constant amplitude and polarization}

As a second example, let us now include a step electromagnetic wave with one
of the impulsive gravitational waves as described in the previous example.
Specifically, let us consider a step electromagnetic wave to come from the
right such that, behind the wavefront, it has constant amplitude and
polarization. This is expressed by the condition
  \begin{equation}
  \label{ConstAmp}
\phi_0=\hbox{const}, \qquad \Psi_0=0,
  \end{equation}
  where the nonzero value of $\phi_0$ determines the constant amplitude and
polarization of the electromagnetic wave, and the vanishing of $\Psi_0$
indicates that there is no associated gravitational wave component behind
the wavefront.

It is first necessary to determine the initial metric functions which give
rise to the specific components (\ref{ConstAmp}). These can be determined
using the expressions (\ref{NPscalars}), together with the
field equation (\ref{EqnAlphaminus}) and the definition of the Ernst
potential (\ref{EFPotentials}). It is found that a general solution of these
equations, which satisfies the junction conditions (\ref{InDataC}) is given
by
  \begin{equation}
  \label{RightWaveE}
\begin{array}{l}
  \alpha_{\scriptscriptstyle-}(v)
=1-(1+k_{\scriptscriptstyle-}^2)\sin^2 (\ell_{\scriptscriptstyle -}v),
  \qquad\qquad f_{\scriptscriptstyle-}(v)=1, \\[2ex]
  {\cal E}_{\scriptscriptstyle-}(v)=-1
-2k_{\scriptscriptstyle-} \sin(\ell_{\scriptscriptstyle-}v)
-2k_{\scriptscriptstyle-}^2
\bigl(1-\cos(\ell_{\scriptscriptstyle-}v)\bigr), \\[2ex]
  \Phi_{\scriptscriptstyle-}(v) =e^{i\gamma_{\scriptscriptstyle-}}
\big[\sin(\ell_{\scriptscriptstyle-}v)+k_{\scriptscriptstyle-}
\bigl(1-\cos(\ell_{\scriptscriptstyle-}v)\bigr) \big], \\[2ex]
H_{\scriptscriptstyle-}(v)=
\big[\cos(\ell_{\scriptscriptstyle-}v) +k_{\scriptscriptstyle-}
\sin(\ell_{\scriptscriptstyle-}v) \big]^2, \qquad\quad
  \Omega_{\scriptscriptstyle-}(v)=0, \\[2ex]
  \widetilde{\Phi}_{\scriptscriptstyle -}(v)
=-ie^{i\gamma_{\scriptscriptstyle-}} \left[\sin(\ell_{\scriptscriptstyle-}v)
-k_{\scriptscriptstyle-}(1-\cos (\ell_{\scriptscriptstyle-}v))\right].
\end{array}
\end{equation}
In the above expressions, $\ell_{\scriptscriptstyle-}$ and
$\gamma_{\scriptscriptstyle-}$ are real constants which determine the values
of the constant amplitude and phase of the electromagnetic field such that
the above scalar is $\phi_0=-2\ell_{\scriptscriptstyle-}
e^{i\gamma_{\scriptscriptstyle-}}$. In addition to these parameters, a
nonzero value of the real constant parameter $k_{\scriptscriptstyle-}$
represents an impulsive gravitational wave localized on the wavefront. If we
want to consider a purely electromagnetic wave, we simply put
$k_{\scriptscriptstyle-}=0$. However, in this subsection we do not use this
simplification, and permit the parameter $k_{\scriptscriptstyle-}$ to be
nonzero.

It is now necessary to obtain the  geometrically defined coordinates in the
interaction region. Using (\ref{Alpha}) with (\ref{NHLwave}) and
(\ref{RightWaveE}) as the characteristic initial data, we obtain
  $$ \alpha(u,v)=1-(1+k_{\scriptscriptstyle -}^2)\sin^2
(\ell_{\scriptscriptstyle -}v)-k_{\scriptscriptstyle +}^2 u^2,\qquad
\beta(u,v)=(1+k_{\scriptscriptstyle-}^2)\sin^2 (\ell_{\scriptscriptstyle-}v)
-k_{\scriptscriptstyle +}^2 u^2. $$
  This leads to the following expressions for $\xi$ and $\eta$:
   \begin{equation}
\label{NullCoordsA} \xi=1-2 k_{\scriptscriptstyle +}^2 u^2, \qquad
\eta=-1+2 (1+k_{\scriptscriptstyle -}^2)\sin^2 (\ell_{\scriptscriptstyle -}
v).
   \end{equation}

\paragraph*{Calculation of the ``in-states'' $\bigpsi_{\scriptscriptstyle
+}(u,w)$ and $\bigpsi_{\scriptscriptstyle -}(v,w)$.}

Using the characteristic initial data (\ref{NHLwave}), (\ref{NHLwaveA}) and
(\ref{RightWaveE}), we can again calculate directly, first the matrix
coefficients $\U(\xi,\eta=-1)$ and $\V(\xi=1,\eta)$, and then the
corresponding fundamental solutions $\bigpsi_{\scriptscriptstyle +}(u,w)$
and $\bigpsi_{\scriptscriptstyle -}(v,w)$ of the linear systems
(\ref{Psipm}) on the characteristics $\eta=-1$ (i.e.
$v=0$) and $\xi=1$ (i.e. $u=0$). However, since the left wave in this case
is the same as that of the first example, the expressions for
$\bigpsi_{\scriptscriptstyle+}(u,w)$ and its inverse are given by
(\ref{NHPsi}) and (\ref{NHPsim1}). The result of similar calculations for
$\bigpsi_{\scriptscriptstyle -}(v,w)$ and its inverse for the right wave is
  $$ \begin{array}{l}
  \bigpsi_{\scriptscriptstyle-}(v,w)=\widetilde{\bpsi}_{0-}(v,w)
\otimes\k_{\scriptscriptstyle-}(w)+ \M_{0-}(v,w),\quad \qquad
  \k_{\scriptscriptstyle -}(w) =
\left(1,\,-i,\,-\displaystyle\frac i{2k_{\scriptscriptstyle-}}
\right), \\[2ex]
  \widetilde{\bpsi}_{0-}(v,w)
  =\lambda_{\scriptscriptstyle-}^{-1}
\left\{\displaystyle\frac{k_{\scriptscriptstyle -}
\sin(\ell_{\scriptscriptstyle -} v)}{w+1}
  \begin{pmatrix}
1\\-i\\0\end{pmatrix}-\displaystyle
\frac{k_{\scriptscriptstyle-}^2\cos(\ell_{\scriptscriptstyle -} v)}{w-w_{1-}}
  \begin{pmatrix} 1\\i\\-4 i k_{\scriptscriptstyle -}\end{pmatrix}+
  \begin{pmatrix} 0\\0\\2 i
k_{\scriptscriptstyle-}\end{pmatrix}\right\},\\[3ex]
  \M_{0-}(v,w)=\displaystyle\frac {k_{\scriptscriptstyle -}^2}{w-w_{1-}}
\begin{pmatrix} 1\\i\\-4 i k_{\scriptscriptstyle -}\end{pmatrix}\otimes
\left(1,\,-i,\,-\displaystyle\frac i{2 k_{\scriptscriptstyle -}}\right)
+\begin{pmatrix} 1&0&0\\ 0&1&0\\
  -2 i k_{\scriptscriptstyle -}&-2 k_{\scriptscriptstyle -}&0\end{pmatrix}
\end{array} $$
  and for the inverse matrix we obtain
  $$ \begin{array}{l}
  \bigpsi_{\scriptscriptstyle -}^{-1}(v,w) =\l_{\scriptscriptstyle-}(w)
\otimes\widetilde{\bphi}_{0-}(v,w)+
\N_{0-}(v,w),\quad\qquad
  \l_{\scriptscriptstyle -}(w)=(w+1)\left(1,\,i,\,-\displaystyle\frac {2 i} {k_{\scriptscriptstyle -}}(w-1)\right)\\[2ex]
  \widetilde{\bphi}_{0-}(v,w)=
  -\displaystyle\frac {\lambda_{\scriptscriptstyle -}
k_{\scriptscriptstyle-}^2}{(w+1)(w-w_{1-})}
\left(1,\,-i,\, -\displaystyle\frac {i}{2 k_{\scriptscriptstyle-}} \right)
\\[3ex]
  \N_{0-}(v,w)=-\displaystyle\frac {k_{\scriptscriptstyle -}\sin
(\ell_{\scriptscriptstyle -} v)}{w+1}
  \begin{pmatrix} 1\\-i\\0\end{pmatrix}\otimes
\left(1,\,-i,\,-\displaystyle\frac i
{2 k_{\scriptscriptstyle -}}\right)\\[3ex]
  \phantom{\N_{0-}(v,w)=}
+\displaystyle\frac {k_{\scriptscriptstyle -}^2\cos
(\ell_{\scriptscriptstyle-} v)}{w-w_{1-}}
\begin{pmatrix} 1\\i\\-4 ik_{\scriptscriptstyle-}
\end{pmatrix}
  \otimes\left(1,\,-i,\,
  -\displaystyle\frac {i}{2 k_{\scriptscriptstyle -}}\right)
+\begin{pmatrix} 1&0&0\\ 0&1&0\\ -2 i k_{\scriptscriptstyle -}
&-2k_{\scriptscriptstyle-} &0 \end{pmatrix}
  \end{array} $$
  where $w_{1-}=1+2 k_{\scriptscriptstyle-}^2$, and the function
$\lambda_{\scriptscriptstyle -}(v,w)$ was defined in (\ref{lambdadef}).

\paragraph*{Construction of the integral evolution equations.}

We now need the conserved monodromy data vectors
   $$ \k_{\scriptscriptstyle+}(w)=\begin{pmatrix}1,& i,\,0\end{pmatrix},
\qquad
  \k_-(w)=\left(1,\,-i,\,-\displaystyle\frac i{2 k_{\scriptscriptstyle-}}
\right) $$
  and the jumps $\bfi_{\scriptscriptstyle 0+}$ and $\bfi_{\scriptscriptstyle
0-}$ of the column-vector functions $\widetilde{\bpsi}_{\scriptscriptstyle
0+}$ at  $\tau_{\scriptscriptstyle +}\in L_{\scriptscriptstyle +}$ and
$\widetilde{\bpsi}_{\scriptscriptstyle 0-}$ at $\tau_{\scriptscriptstyle
-}\in L_{\scriptscriptstyle -}$ respectively:
   $$ \begin{array}{l}
\bfi_{\scriptscriptstyle 0+}(u,\tau_{\scriptscriptstyle+})
=[\lambda_{\scriptscriptstyle +}^{-1}]_{\tau_{\scriptscriptstyle+}}
\left\{\displaystyle\frac 1 2\begin{pmatrix}1\\ -i \\0\end{pmatrix}-
\displaystyle\frac{k_{\scriptscriptstyle +}e^{-i\delta_{\scriptscriptstyle+}}
  u}{\tau_{\scriptscriptstyle +}-1}
\begin{pmatrix}1\\ i \\0\end{pmatrix}\right\}\\[4ex]
  \bfi_{\scriptscriptstyle 0-}(v,\tau_{\scriptscriptstyle -})=
[\lambda_{\scriptscriptstyle -}^{-1}]_{\tau_{\scriptscriptstyle -}}\left\{
\displaystyle\frac{k_{\scriptscriptstyle -}
\sin(\ell_{\scriptscriptstyle -} v)}{\tau_{\scriptscriptstyle -}+1}
  \begin{pmatrix}
1\\-i\\0\end{pmatrix}-\displaystyle\frac{k_{\scriptscriptstyle-}^2
\cos(\ell_{\scriptscriptstyle-}v)} {\tau_{\scriptscriptstyle-}-w_{1-}}
  \begin{pmatrix} 1\\i\\-4 i k_{\scriptscriptstyle -}\end{pmatrix}+
  \begin{pmatrix} 0\\0\\2 i k_{\scriptscriptstyle -}\end{pmatrix}
\right\}\end{array} $$
  which are the characteristic initial values for the unknowns in our integral
evolution equations. Then, we obtain the dynamical monodromy data vectors
using (\ref{DMData}) and the above expressions for the matrix functions
$\bigpsi_{\scriptscriptstyle+}(u,w)$ and
$\bigpsi_{\scriptscriptstyle-}(v,w)$ and their inverse:
  $$ \begin{array}{l}
\m_{\scriptscriptstyle +}(v,\zeta_{\scriptscriptstyle +})
=\Bigl(1,\,i,\,0 \Bigr)-\displaystyle{\frac{ 2 k_{\scriptscriptstyle
-}\sin(\ell_{\scriptscriptstyle -} v)} {\zeta_{\scriptscriptstyle +}+1}}
\Bigl(1,\,-i,\,-\displaystyle\frac i{2 k_{\scriptscriptstyle -}} \Bigr)\\[2ex]
\m_-(u,\zeta_-)=\Bigl(1,\,-i,\,-\displaystyle\frac i{2 
k_{\scriptscriptstyle -}} \Bigr)+
\displaystyle{\frac{2 k_{\scriptscriptstyle +}
e^{-i\delta_{\scriptscriptstyle +}} u}{\zeta_{\scriptscriptstyle 
-}-1}}\Bigl(1,\,i,\,0 \Bigr)
  \end{array}
  $$
   Then, for the kernels $S_\pm$ defined in (\ref{KernelsA}), we 
obtain the expressions
   $$
S_{\scriptscriptstyle +}(u,\tau_{\scriptscriptstyle +},\zeta_-)=
-\displaystyle{\frac{2 k_{\scriptscriptstyle +} 
e^{-i\delta_{\scriptscriptstyle +}} u
[\lambda_{\scriptscriptstyle +}^{-1}]_{\tau_{\scriptscriptstyle +}}}
{i\pi(\zeta_--1)(\tau_{\scriptscriptstyle +}-1)}},\qquad
S_-(v,\tau_-,\zeta_{\scriptscriptstyle +})=
\displaystyle{\frac{2 k_{\scriptscriptstyle -} \sin 
(\ell_{\scriptscriptstyle -} v)
[\lambda_-^{-1}]_{\tau_-}} {i\pi(\zeta_{\scriptscriptstyle +}+1)(\tau_-+1)}}.
 $$
  We now have to solve the integral equations (\ref{IntEqsA}) for the
functions $\bfi_{\scriptscriptstyle+}$ and $\bfi_{\scriptscriptstyle-}$
using the above expressions for the kernels and the characteristic initial
values $\bfi_{\scriptscriptstyle 0+}$ and $\bfi_{\scriptscriptstyle 0-}$.
The procedure involved can be reduced to the solution of an algebraic system
if we observe that the structure of the kernels calculated above and the
right hand sides of these integral equations imply the following dependence
of the unknown functions on the spectral parameter:
  $$  \bfi_{\scriptscriptstyle +}=
  [\lambda_{\scriptscriptstyle +}^{-1}]_{\tau_{\scriptscriptstyle +}} \left\{
  {\bf A}_{\scriptscriptstyle +}
+\displaystyle\frac{{\bf B}_{\scriptscriptstyle +}}
{\tau_{\scriptscriptstyle+}-1},\right\}, \qquad
  \bfi_{\scriptscriptstyle -}=
  [\lambda_{\scriptscriptstyle -}^{-1}]_{\tau_{\scriptscriptstyle -}} \left\{
  {\bf A}_{\scriptscriptstyle -}
+\displaystyle\frac{{\bf B}_{\scriptscriptstyle -}}
{\tau_{\scriptscriptstyle-}+1}
+\displaystyle\frac{{\bf C}_{\scriptscriptstyle -}}
{\tau_{\scriptscriptstyle-} -w_{\scriptscriptstyle 1-}}\right\},
  $$
  where the vector coefficients ${\bf A}_\pm$, ${\bf B}_\pm$ and ${\bf C}_-$
are independent of $\tau_{\scriptscriptstyle\pm}$, although they can be
functions of the coordinates $u$ and $v$. Substituting these expressions
back into the linear integral equations, we obtain that the coefficients
${\bf A}_\pm$ and ${\bf C}_-$ should be expressed as
  $$  {\bf A}_+= \displaystyle\frac 1 2\begin{pmatrix}1\\ -i \\0\end{pmatrix}, \qquad  
{\bf A}_-= \begin{pmatrix} 0\\0\\2 i k_{\scriptscriptstyle-}\end{pmatrix},
\qquad
{\bf C}_-=-k_{\scriptscriptstyle -}^2\cos(\ell_{\scriptscriptstyle -} v) 
  \begin{pmatrix} 1\\i\\-4 i k_{\scriptscriptstyle -}\end{pmatrix}
  $$
  whereas the other coefficients should satisfy algebraic systems whose
coefficients are explicitly calculated integrals and whose solutions are
  $$\begin{array}{l}
  {\bf B}_+ =-\displaystyle\frac{k_{\scriptscriptstyle+}
e^{-i\delta_{\scriptscriptstyle+}} u} {Z(u,v)}
  \begin{pmatrix}
  \widetilde{u}\cos(\ell_{\scriptscriptstyle-}v) +k_{\scriptscriptstyle-}
\sin(\ell_{\scriptscriptstyle-}v) \\[1ex]
  i \widetilde{u}\cos(\ell_{\scriptscriptstyle-}v) -ik_{\scriptscriptstyle-}
\sin(\ell_{\scriptscriptstyle-}v) \\[1ex]
  4 i k_{\scriptscriptstyle -}\widetilde{u}\bigl(1-\cos
(\ell_{\scriptscriptstyle -} v)\bigr)
\end{pmatrix}\\[4ex]
  {\bf B}_-=\displaystyle\frac{k_{\scriptscriptstyle -}
\sin(\ell_{\scriptscriptstyle -} v)} {Z(u,v)}
\begin{pmatrix}
k_{\scriptscriptstyle+} e^{-i\delta_{\scriptscriptstyle+}}u
\cos(\ell_{\scriptscriptstyle-}v)+ \widetilde{v}\\[1ex]
  i k_{\scriptscriptstyle +} e^{-i\delta_{\scriptscriptstyle+}}u
\cos(\ell_{\scriptscriptstyle-}v)- i\widetilde{v}\\[1ex]
  4 i k_{\scriptscriptstyle +} k_{\scriptscriptstyle-}
e^{-i\delta_{\scriptscriptstyle+}}
u\bigl(1-\cos(\ell_{\scriptscriptstyle-}v)\bigr)
\end{pmatrix}
  \end{array}
  $$
  where we have used the notations
$\widetilde{u}=\sqrt{1-k_{\scriptscriptstyle +}^2 u^2}$,
$\widetilde{v}=\sqrt{1-(1+k_{\scriptscriptstyle +}^2)\sin^2
(\ell_{\scriptscriptstyle -} v)}$, and
$Z(u,v)=\widetilde{u}\widetilde{v} -k_{\scriptscriptstyle+}
k_{\scriptscriptstyle-} e^{-i\delta_{\scriptscriptstyle +}}u
\sin(\ell_{\scriptscriptstyle-}v)$. This solution of the integral evolution
equations, in accordance with (\ref{ErnstPotentialsB}), leads to the
following expressions for the Ernst potentials:
  $$\begin{array}{l}
  {\cal E}(u,v)=-1-k_{\scriptscriptstyle +}^2 u^2- 2
k_{\scriptscriptstyle-}^2 \bigl(1-\cos(\ell_{\scriptscriptstyle -} v)\bigr)
-\displaystyle\frac{2 k_{\scriptscriptstyle -}
\sin(\ell_{\scriptscriptstyle-}v)} {Z(u,v)} (\widetilde{v}+
  e^{-i\delta_{\scriptscriptstyle +}}k_{\scriptscriptstyle +}u)\\[2ex]
  \phantom{{\cal E}(u,v)=}
  -\displaystyle\frac{2 e^{-i\delta_{\scriptscriptstyle+}}
k_{\scriptscriptstyle +}u \cos(\ell_{\scriptscriptstyle -} v)}{Z(u,v)}
  (\widetilde{u}+k_{\scriptscriptstyle-} \sin(\ell_{\scriptscriptstyle-}v))
\\[2ex]
  \Phi(u,v)= k_{\scriptscriptstyle -}(1-\cos(\ell_{\scriptscriptstyle -} v))+
  \displaystyle\frac{(\widetilde{v}+
  e^{-i\delta_{\scriptscriptstyle +}}k_{\scriptscriptstyle +}u
  \cos(\ell_{\scriptscriptstyle -} v))
  \sin(\ell_{\scriptscriptstyle -} v)}{Z(u,v)}
  \end{array} $$
  This is a new family of solutions which, for particular values of its
parameters, reduces to various known solutions. Specifically, for pure
vacuum limits, if we substitute  $k_{\scriptscriptstyle-}\to
k_{\scriptscriptstyle-}/\ell_{\scriptscriptstyle-}$ and then take the limit
$\ell_{\scriptscriptstyle-}\to 0$, we obtain the Khan--Penrose solution
\cite{Khan-Penrose:1971} if $\delta_{\scriptscriptstyle+}=0$ or the
Nutku--Halil solution \cite{Nutku-Halil:1977} if
$\delta_{\scriptscriptstyle+}\ne 0$ as in the previous example. The limit
$k_{\scriptscriptstyle -}=0$, $\delta_{\scriptscriptstyle +}=0$ leads to a
known solution \cite{Griffiths:1991} for the collision of a step
electromagnetic wave with an impulsive gravitational wave. The case
$k_{\scriptscriptstyle-}\ne 0$, $\delta_{\scriptscriptstyle+}=0$ corresponds
to a generalization of the solution of Hogan, Barrab\`es and Bressange
\cite{Hogan-Barrabes-Bressange:1998}.

\subsection{Collision of electromagnetic step-like waves of constant
amplitudes and polarizations}

As a final example, let us consider the collision of two step
electromagnetic waves of constant amplitudes and polarizations which are not
accompanied by gravitational impulses. It can be seen from above that, for
this case, we can take
   \begin{equation}\label{BSGWavesA}
   \begin{array}{lcccl}
  \alpha_{\scriptscriptstyle+}(u) =\cos^2 (\ell_{\scriptscriptstyle +}u)
  &&&&\alpha_{\scriptscriptstyle-}(v)
=\cos^2(\ell_{\scriptscriptstyle-}v),\\[2ex]
  {\cal E}_{\scriptscriptstyle+}(u)=-1&&&&
{\cal E}_{\scriptscriptstyle-}(v)=-1\\[2ex]
  \Phi_{\scriptscriptstyle+}(u) =e^{i\gamma_{\scriptscriptstyle+}}
\sin(\ell_{\scriptscriptstyle +}u) &&&&
\Phi_{\scriptscriptstyle -}(v) =e^{i\gamma_{\scriptscriptstyle-}}
\sin(\ell_{\scriptscriptstyle-}v)
\end{array}
  \end{equation}
  with $f_{\scriptscriptstyle+}(u)=1$ and $f_{\scriptscriptstyle-}(v)=1$. The
metric functions and additional component of the complex electromagnetic
potential then take the forms
  \begin{equation}
  \label{BSGWavesB}
\begin{array}{lcccl}
H_{\scriptscriptstyle +}(u)=\cos^2 (\ell_{\scriptscriptstyle +}u),
&&&& H_{\scriptscriptstyle-}(v)=\cos^2 (\ell_{\scriptscriptstyle -}v), \\[1ex]
\Omega_{\scriptscriptstyle+}(u) =0,
&&&&\Omega_{\scriptscriptstyle-}(v) =0),\\[1ex]
  \widetilde{\Phi}_{\scriptscriptstyle+}(u) = i 
e^{i\gamma_{\scriptscriptstyle+}}
\sin(\ell_{\scriptscriptstyle +}u) &&&&
\widetilde{\Phi}_{\scriptscriptstyle -}(v) =-i
e^{i\gamma_{\scriptscriptstyle-}}
\sin(\ell_{\scriptscriptstyle-}v)
\end{array}
  \end{equation}
  where the real parameters $\ell_{\scriptscriptstyle\pm}$ and
$\gamma_{\scriptscriptstyle\pm}$ determine the initial amplitude and phase
respectively of the two approaching electromagnetic waves.

In this case, the geometrically defined coordinates in the interaction
region are
  $$ \begin{array}{l}
  \alpha(u,v)=1-\sin^2 (\ell_{\scriptscriptstyle +}u)-\sin^2
(\ell_{\scriptscriptstyle -}v), \\[1ex]
  \beta(u,v)=\sin^2 (\ell_{\scriptscriptstyle -}v)-\sin^2
(\ell_{\scriptscriptstyle +}u),
  \end{array} $$
  and the corresponding expressions for $\xi$ and $\eta$ are
   \begin{equation} \label{NullCoordsB}
   \xi=1-2\sin^2 (\ell_{\scriptscriptstyle +}u), \qquad
   \eta=-1+2 \sin^2 (\ell_{\scriptscriptstyle -}v).
   \end{equation}

\paragraph*{Calculation of the ``in-states'' $\bigpsi_{\scriptscriptstyle
+}(u,w)$ and $\bigpsi_{\scriptscriptstyle -}(v,w)$}

As in the previous examples, with the characteristic initial data
(\ref{BSGWavesA}), we can calculate, first the matrix coefficients
$\U(\xi,\eta=-1)$ and $\V(\xi=1,\eta)$, and then the corresponding
fundamental solutions $\bigpsi_{\scriptscriptstyle +}(u,w)$ and
$\bigpsi_{\scriptscriptstyle -}(v,w)$ of the linear systems (\ref{Psipm})
on the characteristics $\eta=-1$ (i.e. $v=0$) and
$\xi=1$ (i.e. $u=0$). The result of these calculations for
$\bigpsi_{\scriptscriptstyle +}(u,w)$ and its inverse is
  $$ \begin{array}{l}
\bigpsi_{\scriptscriptstyle+}(u,w)
=\widetilde{\bpsi}_{\scriptscriptstyle0+}(u,w)
\otimes\k_{\scriptscriptstyle+}(w)+
\M_{\scriptscriptstyle 0+}(u,w),\qquad 
\k_{\scriptscriptstyle +}(w)=\left(0,\,0,\,1\right),\\[2ex]
  \widetilde{\bpsi}_{\scriptscriptstyle 0+}(u,w)
=\lambda_{\scriptscriptstyle+}^{-1}(u,w) \begin{pmatrix}
-\displaystyle\frac{i e^{i\gamma_{\scriptscriptstyle+}}
\sin(\ell_{\scriptscriptstyle+}u)} {2(w-1)}\\
  \displaystyle\frac{e^{i\gamma_{\scriptscriptstyle+}}
\sin(\ell_{\scriptscriptstyle+}u)} {2(w-1)} \\ 1\end{pmatrix},
  \qquad
  \M_{\scriptscriptstyle 0+}(u,w)=\begin{pmatrix} 1&0&0\\ 0&1&0\\
0&0&0\end{pmatrix}
\end{array}
  $$
  where $\lambda_{\scriptscriptstyle +}(u,w)$ is defined in
(\ref{lambdadef}). The inverse matrices possess the expressions
$$\begin{array}{l}
\bigpsi_{\scriptscriptstyle
+}^{-1}(u,w)=\l_{\scriptscriptstyle
+}(w)\otimes\widetilde{\bphi}_{\scriptscriptstyle 0+}(u,w)+
\N_{\scriptscriptstyle 0+}(u,w),\qquad
\l_{\scriptscriptstyle +}(w)=-4(w^2-1)\begin{pmatrix}
0\\0\\1\end{pmatrix}\\[4ex]
\widetilde{\bphi}_{\scriptscriptstyle 0+}(u,w)=
-\displaystyle\frac{\lambda_{\scriptscriptstyle +}(u,w)}
{4(w^2-1)}
\left(0,\,0,\, 1 \right),\qquad
\N_{0+}(u,w)=\begin{pmatrix} 1&0& \displaystyle\frac{i
e^{i\gamma_{\scriptscriptstyle
+}}\sin(\ell_{\scriptscriptstyle +} u)}{2(w-1)}\\
0&1&-\displaystyle\frac{e^{i\gamma_{\scriptscriptstyle
+}}\sin(\ell_{\scriptscriptstyle +} u)}{2(w-1)}\\
0&0&0\end{pmatrix}
\end{array}
$$
  The equivalent expressions for $\bigpsi_{\scriptscriptstyle-}(v,w)$ and
its inverse have the same structure except that
  $$ \widetilde{\bpsi}_{\scriptscriptstyle 0-}(v,w) =\lambda_{\scriptscriptstyle
-}^{-1}(v,w)\begin{pmatrix} -\displaystyle\frac{i 
e^{i\gamma_{\scriptscriptstyle -}}\sin(\ell_{\scriptscriptstyle -} 
v)}{2(w+1)}\\
-\displaystyle\frac{e^{i\gamma_{\scriptscriptstyle 
-}}\sin(\ell_{\scriptscriptstyle -} v)}{2(w+1)}\\
1\end{pmatrix},\qquad
\N_{0-}(v,w)=\begin{pmatrix} 1&0& \displaystyle\frac{i
e^{i\gamma_{\scriptscriptstyle
-}}\sin(\ell_{\scriptscriptstyle -} v)}{2(w+1)}\\
0&1&\displaystyle\frac{e^{i\gamma_{\scriptscriptstyle
-}}\sin(\ell_{\scriptscriptstyle -} v)}{2(w+1)}\\
0&0&0\end{pmatrix}. $$

\paragraph*{Construction of the integral evolution equations.}

We now need certain fragments of the structures of the matrix functions
$\bigpsi_{\scriptscriptstyle+}(u,w)$ and
$\bigpsi_{\scriptscriptstyle-}(v,w)$ on the cuts $L_\pm$. These are the
conserved monodromy data vectors
   $$ \k_{\scriptscriptstyle +}(w)=\begin{pmatrix}0,& 0,\,1\end{pmatrix},\qquad
   \k_-(w)=\left(0,\,0,\,1\right) $$
and the jumps $\bfi_{\scriptscriptstyle 0+}$ and
$\bfi_{\scriptscriptstyle 0-}$ of the column-vector
functions $\widetilde{\bpsi}_{\scriptscriptstyle 0+}$ at
$\tau_{\scriptscriptstyle +}\in L_{\scriptscriptstyle +}$
and $\widetilde{\bpsi}_{\scriptscriptstyle 0-}$ at
$\tau_{\scriptscriptstyle -}\in L_{\scriptscriptstyle -}$
respectively:
   $$
\bfi_{\scriptscriptstyle 0+}(u,\tau_{\scriptscriptstyle
+})=[\lambda_{\scriptscriptstyle
+}^{-1}]_{\tau_{\scriptscriptstyle +}}
\begin{pmatrix} -\displaystyle\frac{i e^{i\gamma_{\scriptscriptstyle +}}
\sin(\ell_{\scriptscriptstyle +} u)}{2(\tau_{\scriptscriptstyle +}-1)}\\
\displaystyle\frac{e^{i\gamma_{\scriptscriptstyle +}}
\sin(\ell_{\scriptscriptstyle +} u)}{2(\tau_{\scriptscriptstyle +}-1)}\\
1\end{pmatrix},\qquad
\bfi_{\scriptscriptstyle
0-}(v,\tau_{\scriptscriptstyle -})=
[\lambda_{\scriptscriptstyle
-}^{-1}]_{\tau_{\scriptscriptstyle -}}
\begin{pmatrix} -\displaystyle\frac{i e^{i\gamma_{\scriptscriptstyle -}}
\sin(\ell_{\scriptscriptstyle -} v)}{2(\tau_{\scriptscriptstyle +}+1)}\\
-\displaystyle\frac{e^{i\gamma_{\scriptscriptstyle -}}
\sin(\ell_{\scriptscriptstyle -} v)}{2(\tau_{\scriptscriptstyle +}+1)}\\
1\end{pmatrix},
$$
which are the characteristic initial values for the unknowns
in our integral evolution equations. After that, we
calculate the dynamical monodromy data vectors
  $$
\m_{\scriptscriptstyle +}(v,\zeta_{\scriptscriptstyle
+})=\lambda_{\scriptscriptstyle -}
(\zeta_{\scriptscriptstyle +}) \Bigl(0,\,0,\,1
\Bigr),\qquad
\m_{\scriptscriptstyle -}(u,\zeta_{\scriptscriptstyle
-})=\lambda_{\scriptscriptstyle +}
(\zeta_{\scriptscriptstyle -}) \Bigl(0,\,0,\,1 \Bigr).
  $$
Then, for the kernels $S_\pm$ defined in (\ref{KernelsA}),
we obtain the expressions
   $$
S_{\scriptscriptstyle +}(u,\tau_{\scriptscriptstyle
+},\zeta_-)= \displaystyle{\frac{
\lambda_{\scriptscriptstyle +}(u,\zeta_{\scriptscriptstyle
-})}
{i\pi(\zeta_{\scriptscriptstyle -}-\tau_{\scriptscriptstyle
+})}},\qquad
S_{\scriptscriptstyle -}(v,\tau_{\scriptscriptstyle
-},\zeta_{\scriptscriptstyle +})= \displaystyle{\frac{
\lambda_{\scriptscriptstyle -}(v,\zeta_{\scriptscriptstyle
+})} {i\pi(\zeta_{\scriptscriptstyle
+}-\tau_{\scriptscriptstyle -})}}
$$
With these expressions, the structure of the integral
evolution equations (\ref{IntEqsA}) implies the following
dependence of the unknown functions on the spectral
parameter:
  $$\begin{array}{l}
  \bfi_{\scriptscriptstyle +}(\tau_{\scriptscriptstyle +})=
  [\lambda_{\scriptscriptstyle +}^{-1}]_{\tau_{\scriptscriptstyle +}}
\lambda_{\scriptscriptstyle -
}^{-1}(\tau_{\scriptscriptstyle +}) \left\{
\begin{pmatrix} 0\\0\\1\end{pmatrix}+
\displaystyle\frac{{\bf A}_{\scriptscriptstyle+}}
{\tau_{\scriptscriptstyle+}-1}+ \displaystyle\frac{{\bf
B}_{\scriptscriptstyle +}} {\tau_{\scriptscriptstyle+}+1}\right\}\\[4ex]
  \bfi_{\scriptscriptstyle -}(\tau_{\scriptscriptstyle -})=
  [\lambda_{\scriptscriptstyle -}^{-1}]_{\tau_{\scriptscriptstyle -}}
\lambda_{\scriptscriptstyle +}^{-1}(\tau_{\scriptscriptstyle-}) \left\{
\begin{pmatrix} 0\\0\\1\end{pmatrix}+
\displaystyle\frac{{\bf A}_{\scriptscriptstyle-}}
{\tau_{\scriptscriptstyle-}-1}+ \displaystyle\frac{{\bf
B}_{\scriptscriptstyle -}}{\tau_{\scriptscriptstyle-}+1}\right\}
  \end{array} $$
  where the vector coefficients ${\bf A}_\pm$ and ${\bf B}_\pm$ are
independent of $\tau_{\scriptscriptstyle\pm}$ but can be functions of $u$
and $v$. Substituting these expressions back into the linear integral
equations, we obtain
  $$ 
  {\bf A}_+=  {\bf A}_-=-\displaystyle\frac i 2
e^{i\gamma_{\scriptscriptstyle+}}
  \sin(\ell_{\scriptscriptstyle +} u)\cos(\ell_{\scriptscriptstyle -} v)
  \begin{pmatrix}1\\i \\0\end{pmatrix},\quad
{\bf B}_+=  {\bf B}_-=-\displaystyle\frac i 2
e^{i\gamma_{\scriptscriptstyle -}}
  \cos(\ell_{\scriptscriptstyle +} u)\sin(\ell_{\scriptscriptstyle -} v)
  \begin{pmatrix}1\\-i \\0\end{pmatrix}
 $$
  This solution of the integral evolution equations, in accordance with
(\ref{ErnstPotentialsB}) leads to the following expressions for the Ernst
potentials:
  $${\cal E}=-1,\qquad
\Phi= e^{i\gamma_{\scriptscriptstyle +}}\sin(\ell_{\scriptscriptstyle 
+} u)\cos(\ell_{\scriptscriptstyle -} v) + 
e^{i\gamma_{\scriptscriptstyle -}}\cos(\ell_{\scriptscriptstyle +} 
u)\sin(\ell_{\scriptscriptstyle -} v).
$$
  In the particular case in which the electromagnetic waves possess aligned
polarizations (i.e. for $\gamma_{\scriptscriptstyle+}
=\gamma_{\scriptscriptstyle-}$), this solution reduces to the Bell--Szekeres
solution while, for $\gamma_{\scriptscriptstyle+}
\ne\gamma_{\scriptscriptstyle-}$, it coincides with another
solution of Griffiths (see
\cite{Griffiths:1991}) with the metric functions
  $$ \begin{array}{l}
  \alpha(u,v)=1-\sin^2 (\ell_{\scriptscriptstyle +}u)-\sin^2
(\ell_{\scriptscriptstyle -}v),\\[2ex]
  H(u,v)=\cos^2(\ell_{\scriptscriptstyle+}u)
\cos^2(\ell_{\scriptscriptstyle-} v)+\cos^2(\ell_{\scriptscriptstyle+}u)
\cos^2(\ell_{\scriptscriptstyle-}v) -\frac12
\cos(\gamma_{\scriptscriptstyle+} -\gamma_{\scriptscriptstyle-})
\sin(2\ell_{\scriptscriptstyle+}u)
\sin(2\ell_{\scriptscriptstyle-}v),\\[2ex]
  \Omega(u,v)=\displaystyle\frac {\sin(\gamma_{\scriptscriptstyle+}
-\gamma_{\scriptscriptstyle-}) \sin(2\ell_{\scriptscriptstyle+}u)
\sin(2\ell_{\scriptscriptstyle-}v)} {2 H(u,v)}.
  \end{array} $$

\section{Concluding remarks}

We have presented in this paper a general scheme for the construction of
solutions of the characteristic initial value problem for the Einstein or
Einstein--Maxwell field equations for vacuum or electrovacuum space-times
respectively with two-dimensional spatial symmetries. Specifically, we have demonstrated the application of this technique in the construction of solutions of the characteristic initial value problem for the collision and subsequent interaction of given plane gravitational or gravitational and electromagnetic waves with distinct wavefronts, which are propagating initially towards each other in a Minkowski background.

The main feature of this approach is that it enables a solution to be
constructed starting directly from given characteristic initial data. In the
case of plane wave collisions, these data can be determined via the matching
conditions from certain functions and parameters that are specified on the
two intersecting null characteristics which correspond to the wavefronts of
the approaching waves. Then, to construct the corresponding solution in the
wave interaction region, it is necessary to work through three essential
steps.
  \begin{itemize}
  \item{First, we must solve two systems of linear ordinary differential
equations with a spectral parameter which determine the characteristic
initial values ${\bf \Psi}_+(u,w)$ and ${\bf \Psi}_-(v,w)$ of the matrix
function ${\bf \Psi}(u,v,w)$. }
  \item{Next, the solutions ${\bf \Psi}_+(u,w)$ and ${\bf \Psi}_-(v,w)$
should be used in the construction of the kernels and right hand sides of
the integral evolution equations derived above, and then we have to solve
these integral equations.}
  \item{As the last step of our construction, we have to calculate the
quadratures which express all components of the solution of the
characteristic initial value problem in terms of the solution of the
integral evolution equations.}
\end{itemize}

It is clear, that these three steps solve our (effectively two-dimensional)
characteristic initial value problem, at least in principle. However, it is
also clear that a practical realization of all these steps, for more or less
nontrivial initial data, may meet a number of technical difficulties. First
of all, a problem which frequently arises is that, for given initial data,
the ordinary linear differential equations for ${\bf \Psi}_+(u,w)$ and (or)
${\bf \Psi}_-(v,w)$ may not possess explicit solutions. The algorithm
obviously fails in this case. However, even if we are successful in
obtaining an explicit solution of these equations, we may meet a similar difficulty in the solution of the corresponding integral evolution
equations. 

Despite the above-mentioned difficulties, however, the algorithm presented here can work in practice, as is demonstrated in the examples given above. Of course, as may be noted with a grain of salt, most of our examples correspond to physically important but well known solutions. Thus, it has not been demonstrated in the paper that the suggested algorithm can work effectively as a new explicit solution generating (or more correctly, -- solution constructing) technique, although the general solution given in section 7.2 is new. It may be noted however, that our examples were selected using the simplest physically significant structures of the approaching waves. Some further physically reasonable but more complicated examples lead to differential and integral equations which seem not to admit explicit solutions in terms of elementary functions. In these cases further work may be useful. On the other hand, it is not difficult now to find many formally new examples for specially constructed, but more artificial, explicitly ``integrable'' initial data. However, looking for such examples is not of interest for us because it returns to an approach which is opposite to our strategy of construction of solutions {\em starting from the initial data}.

It may thus occur that our approach can be useful for a direct construction of some new physically interesting (exact or possibly approximate) examples. However, it must be pointed out that the main purpose of our construction is a systematic description of the steps which are necessary to actually construct the solution from the initial data. In addition, this approach to the solution of the characteristic initial value problems for vacuum and electrovacuum fields in space-times with two commuting isometries suggests some general frameworks for a consideration of various questions concerning the dynamics of these fields. Among these, we can mention for example, the relation between the structure of the initial waves and the asymptotic behaviour of the solution in the wave interaction region near the singularity, the conditions for the approaching waves leading to the creation of an (unstable) horizon instead of a singularity etc. It may also be mentioned that the basic constructions of the approach suggested above, such as the spectral problems which give rise to the characterisation of every solution in terms of its monodromy data and the linear integral evolution equations, can be used for a similar consideration (and possibly for a solution) of the analogous problems for the collision and interaction of non-plane waves or various waves propagating through non-Minkowskian backgrounds. All these questions seem to be very interesting topics for further investigations.

\section{Acknowledgments}

The work of GAA is supported partly by the Russian Foundation for Basic Research (the grants 02-01-00729 and 02-02-17372) and by the programs ``Nonlinear dynamics'' of Russian Academy of Sciences and ``Leading Scientific Schools'' of Russian Federation (the grant NSh-1697.2003.1).

\end{document}